\documentclass[aps,pre,reprint,superscriptaddress,amsfonts,amsmath,showpacs,nofootinbib,floatfix]{revtex4-1}

\usepackage{xcolor}
\definecolor{darkgreen}{rgb}{0,0.6,0}
\definecolor{cyan}{rgb}{0,0.7,0.8}
\usepackage[colorlinks=true,citecolor=darkgreen,urlcolor=cyan]{hyperref}
\usepackage{graphicx}
\usepackage{multirow}
\usepackage{enumitem}
\setlist{noitemsep}
\usepackage{bm}

\newcommand{\mrm}[1]{\mathrm{#1}}

\newcommand{\mbb}[1]{\mathbb{#1}}
\newcommand{\mc}[1]{\mathcal{#1}}

\newcommand{\eref}[1]{(\ref{#1})}
\newcommand{\Eref}[1]{Eq.~(\ref{#1})}
\newcommand{\Erefr}[2]{Eqs.~(\ref{#1})--(\ref{#2})}
\newcommand{\Erefs}[2]{Eqs.~(\ref{#1}) and (\ref{#2})}
\newcommand{\fref}[1]{Fig.~\ref{#1}}
\newcommand{\tref}[1]{Table~\ref{#1}}
\newcommand{\sref}[1]{Sec.~\ref{#1}}
\newcommand{\scref}[1]{scenario~\ref{#1}}
\newcommand{\pscref}[1]{\protect{scenario~\ref{#1}}}

\newcommand{\aref}[1]{Appendix~\ref{#1}}

\newcommand{\pEref}[1]{\protect{Eq.~(\ref{#1})}}
\newcommand{\pfref}[1]{\protect{Fig.~\ref{#1}}}
\newcommand{\ptref}[1]{\protect{Table~\ref{#1}}}
\newcommand{\psref}[1]{\protect{Sec.~\ref{#1}}}

\newcommand{\ra}{\rangle}

\newcommand{\p}[1]{\phantom{#1}}
\newcommand{\rcite}[1]{Ref.~\onlinecite{#1}}
\newcommand{\prcite}[1]{Ref.~\protect{\onlinecite{#1}}}
\newcommand{\pcite}[1]{\protect{\cite{#1}}}

\newcommand{\MATLAB}{\textsc{Matlab}}
\newcommand{\ttt}[1]{\texttt{#1}}
\newcommand{\OO}[1]{\mrm{O}(#1)}

\newcommand{\xicap}{{\mu_\mrm{cap}}}

\newcommand{\prtext}[1]{}
\newcommand{\arXtext}[1]{#1}
\newcommand{\xa}{\xi_a}
\newcommand{\xb}{\xi_b}
\newcommand{\xc}{\xi_c}
\newcommand{\xd}{\xi_d}
\newcommand{\xe}{\xi_e}

\newcommand{\fuse}[2]{(#1#2)} %
\newcommand{\varsubsection}[1]{\subsection{#1}}
\newcommand{\srefp}[1]{\sref{#1}}

\newcommand{\refp}[1]{\ref{#1}}
\newcommand{\Erefp}[1]{\Eref{#1}}
\newcommand{\condaref}[1]{\aref{#1}}

\newtheorem{lemma}{Lemma}

\begin{document}

\title{Faster identification of optimal contraction sequences for tensor networks}

\author{Robert N. C. Pfeifer}
\email[]{robert.pfeifer@mq.edu.au}
\affiliation{Perimeter Institute for Theoretical Physics, 31 Caroline St. N, Waterloo ON~~N2L 2Y5, Canada}
\author{Jutho Haegeman}
\affiliation{Department of Physics and Astronomy, Ghent University, Krijgslaan 281-S9, B-9000 Ghent, Belgium}
\author{Frank Verstraete}
\affiliation{Department of Physics and Astronomy, Ghent University, Krijgslaan 281-S9, B-9000 Ghent, Belgium}
\affiliation{Vienna Center for Quantum Science, Universit\"{a}t Wien, Boltzmanngasse 5, A-1090 Wien, Austria}

\date{\today}

\begin{abstract}
The efficient evaluation of tensor expressions involving sums over multiple indices is of significant importance to many fields of research, including quantum many-body physics, loop quantum gravity, and quantum chemistry. The computational cost of evaluating an expression may depend strongly upon the order in which the index sums are evaluated, and determination of the operation-minimising contraction sequence for a single tensor network (single term, in quantum chemistry) is known to be NP-hard. The current preferred solution is an exhaustive search, using either an iterative depth-first approach with pruning or dynamic programming and memoisation, but these approaches are impractical for many of the larger tensor network Ans\"atze encountered in quantum many-body physics. We present a modified search algorithm with enhanced pruning which exhibits a performance increase of several orders of magnitude while still guaranteeing identification of an optimal operation-minimising contraction sequence for a single tensor network. A reference implementation for \MATLAB{}, compatible with the \ttt{ncon()} and \ttt{multienv()} network contractors of \href{http://arxiv.org/abs/1402.0939}{arXiv:1402.0939} and \href{http://arxiv.org/abs/1310.8023}{arXiv:1310.8023} respectively, is supplied.
\end{abstract}


\maketitle

\section{Introduction}

\subsection{Overview}

The need to efficiently contract a tensor expression is one which arises in many different areas of research, including both quantum chemistry and quantum many-body physics, and has been an area of intense study since at least 1997 \cite[see e.g.][]{auer2006,baumgartner2002,baumgartner2002a,baumgartner2005,cociorva2001,cociorva2001a,cociorva2002,cociorva2002a,cociorva2003,hartono2005,hartono2006,hirata2003,hirata2004,lam1997,lam1997a,lam1999,lam2000}, despite being known to be NP-hard \cite{lam1997}. %
In quantum many-body physics interest in this problem has been driven by the increasing complexity of tensor network Ans\"atze for many-body systems (see \sref{sec:TNAs}), while in quantum physical chemistry it has long been acknowledged that determination of the optimal contraction sequences for increasingly complex tensor networks represents a significant bottleneck in the development of new algorithms \cite{baumgartner2002,hirata2004,baumgartner2005,auer2006}. %

In its most general form, the problem is to evaluate a multidimensional tensor sum such as
\begin{equation}
\sum_{j,k,m,p}A_{ijk}B_{jlm}C_{kmp}D_{pq}+\sum_{j,k}A_{ijk}D_{jklq}
\end{equation}
as rapidly as possible subject to the constraints of available computing hardware. This problem may be seen as a generalisation of the matrix-chain multiplication problem, described in \aref{sec:matmult}, where a string of matrices are to be multiplied together as efficiently as possible. Unlike the matrix-chain multiplication problem, however, this problem cannot be solved in polynomial time through the use of dynamic programming techniques \cite{lam1997}.

While this optimisation problem is intrinsically multidimensional, balancing available memory and (for multi-node machines) inter-node communication delays against the number of floating-point operations which must be performed, the predominant approach to this problem is first to identify the ideal contraction procedure which would be performed on a single node with infinite resources, 
minimising the number of floating point operations to be performed (\emph{operation minimisation}),
before trading off performance against memory constraints and distributing the problem across multiple nodes \cite{baumgartner2002,cociorva2002,cociorva2002a,hirata2003,lam1997a,lam2000}. Consequently, the task of operation minimisation is of fundamental importance.

A significant difference between applications in quantum chemistry and quantum many-body physics is that the Ans\"atze of the former frequently yield tensor expressions involving many summed terms which may be factorised in numerous different ways, and it is necessary to explore both the different factorisations and the different index contraction sequences for a given factorisation, with the latter task having been termed \emph{single term optimisation} \cite{hartono2005}. On the other hand, in quantum many-body physics an Ansatz is typically made up of a single term, but this term may involve substantially more factors than are encountered in the single terms of quantum chemistry. 
The problem of factorising multiple-term tensor expressions for optimal computational efficiency has been explored in depth elsewhere \cite{hartono2005,hartono2006}, and itself depends on the determination of efficient contraction sequences for single terms, so the task of single term operation minimisation is of key importance in both quantum chemistry and quantum many-body physics (where it is synonymous with \emph{optimal contraction of a tensor network}). It is the problem of rapid single term optimisation which is addressed in this paper.

As the primary interest of the authors is in quantum many-body physics, we will frequently favour quantum physics terminology over that of quantum chemistry or linear algebra. It should therefore be noted that a \emph{tensor network} is synonymous with a \emph{single term in a tensor expression}, and that the \emph{optimal} contraction sequence is assumed to be that which minimises the number of floating point operations performed. For a single term, if it is assumed that evaluation of
\begin{equation}
C_{ij}=\sum_k A_{ik}\times B_{kj}
\end{equation}
begins by constructing an array of appropriate size for matrix $C_{ij}$ and populated with zeros, then minimisation of the number of floating point operations is synonymous with minimising the number of floating point multiplications as the ratio of multiplications to additions is 1:1. We note that identifying an optimal contraction sequence for a single tensor network corresponds to solving the ``multiplication problem'' of \rcite{lam1997} and thus is known to be NP-hard.

An index $i$ runs from 1 to $|i|$, where $|i|$ is termed the \emph{dimension} of index $i$ and may also be denoted $\xi_i$. The dimension of a tensor, e.g. $|C_{ijk}|$, is then the product of the dimension of its indices, and corresponds to the number of entries in the multidimensional array $C_{ijk}$. 

\begin{figure}
\includegraphics[width=246.0pt]{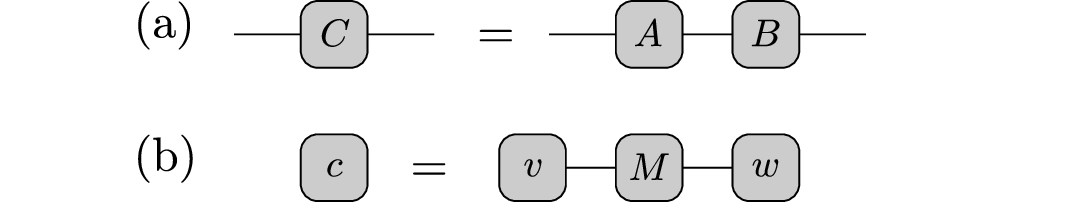}
\caption{Example graphical representations of tensor networks: (a)~Matrix multiplication, $C_{ij}=A_{ik}B^k_{~j}$, and (b)~inner product of a matrix and two vectors, $c=v^\mrm{T}Mw$.\label{fig:examplenetworks}}
\end{figure}

For complicated tensor networks involving large numbers of tensors and index sums, we will use a graphical notation which is a simplification of that of \citeauthor{penrose1971a} \cite{penrose1971a} and is summarised in \S{}1.2 of \rcite{pfeifer2011a}. In this notation, shapes represent tensors and lines (or \emph{legs}) represent indices. A line connecting two shapes therefore represents an index appearing on the tensors corresponding to both shapes. We follow the Einstein summation convention, where any repeated index appearing once in the upper position and once in the lower position is assumed summed, unless otherwise specified. The multiplication of two matrices $C=A\times B$ may therefore be written
\begin{equation}
C_{ij} = A_{ik}B^k_{\p{k}j}\label{eq:AtimesB}
\end{equation}
and represented graphically as in \fref{fig:examplenetworks}(i), while the inner product of a matrix and two column vectors, $v^\mrm{T}Mw$, may be written
\begin{equation}
c = v_i M^{ij} w_j\label{eq:vMw}
\end{equation}
and represented graphically as in \fref{fig:examplenetworks}(ii).

\subsection{Tensor network algorithms in quantum many-body physics\label{sec:TNAs}}

Tensor network algorithms provide powerful tools for the study of a wide variety of physical systems. They are perhaps best known for their use in condensed matter physics as numerical techniques for the study of quantum many-body systems on a lattice \cite[e.g.][]{feiguin2007,trebst2008,konig2010,pfeifer2009,pfeifer2010,cincio2008,evenbly2009,evenbly2009b,evenbly2010a,noack1993,fannes1992,friedman1997,otsuka1996,tagliacozzo2009,murg2007,murg2009,muth2010,murg2010,montangero2009,corboz2010,corboz2010a,corboz2012,corboz2011,yan2011,depenbrock2012}, %
but recent 
breakthroughs 
blending ideas from quantum information with advanced numerical techniques 
have led to the construction of new Ans\"atze \cite[e.g.][]{vidal2007,vidal2008a,evenbly2009,evenbly2009b,vidal2010,evenbly2011} having applications in fields as diverse as holography and the AdS/CFT correspondence \cite{swingle2012,swingle2012a,singh2013}, the study of many-body entanglement \cite{vidal2007,tagliacozzo2009,evenbly2011,evenbly2014a}, and the classification of topological phases in quantum spin systems \cite{chen2011,chen2011a,schuch2011,cincio2013}. As a numerical tool, a tensor network algorithm typically comprises an Ansatz for the description of pure or mixed quantum states, which is composed of a network of tensors, and an iterative procedure for updating this Ansatz. %
Examples include the Density Matrix Renormalisation Group (DMRG) \cite{white1992,schollwock2011} and Time Evolving Block Decimation (TEBD) algorithms \cite{vidal2004,vidal2007b}, both of which are based on the Matrix Product State (MPS) 
Ansatz, and also Tree Tensor Networks (TTNs) \cite{shi2006}, Projected Entangled Pair States (PEPS) \cite{verstraete2004,jordan2008}, and the Multi-scale Entanglement Renormalisation Ansatz (MERA) \cite{vidal2007,vidal2008a,vidal2010,evenbly2009,evenbly2009b,evenbly2011,evenbly2014b}. 

The fundamental challenge to the numerical study of quantum many-body systems on a lattice is that the number of degrees of freedom, and thus the computational cost associated with exact simulation, grows exponentially with the size of the system. To overcome this challenge, tensor network Ans\"atze replace the coefficients $c_{i_1\ldots i_n}$ of a quantum state $|\psi\ra$
\begin{equation}
|\psi\ra=\sum_{i_1\ldots i_n}c_{i_1\ldots i_n}|i_1,\ldots,i_n\ra\label{eq:state}
\end{equation}
with a network of tensors whose dimensions are such that the number of coefficients required to describe the network exhibits a better scaling in $n$, the number of lattice sites, than does the number of coefficients $c_{i_1\ldots i_n}$ in \Eref{eq:state}. Indeed, for many tensor networks this scaling in $n$ is polynomial rather than exponential.

Given this reduction in the number of coefficients in the description, a tensor network Ansatz is capable only of representing states which lie within some restricted region of the Hilbert space of the system, but nevertheless these Ans\"atze and associated algorithms are capable of providing substantial insight into the physics of a wide variety of %
systems in appropriate regimes (e.g.~1D quantum critical systems, systems with limited long-range entanglement, 2D systems obeying an area law. Details vary with the specifics of the tensor network employed---%
see Refs.~\onlinecite
{feiguin2007,trebst2008,konig2010,pfeifer2009,pfeifer2010,cincio2008,evenbly2009,evenbly2009b,evenbly2010a,noack1993,fannes1992,friedman1997,otsuka1996,tagliacozzo2009,murg2007,murg2009,muth2010,murg2010,montangero2009,corboz2010,corboz2010a,corboz2012,corboz2011,yan2011,depenbrock2012} for examples)%
.

In order for a tensor network algorithm to be useful as a tool for numerical computation, it must be possible to perform the operations of the algorithm for a reasonable computational cost. An economical description of the relevant part of the Hilbert space of a system is a good start, but this is not the only factor which must be taken into account: When determining whether a given tensor network algorithm is computationally feasible, the structure of the tensor network itself also plays a significant role. In describing scaling of the cost of a tensor network algorithm, it is customary to express this in the form of a polynomial in some refinement parameter $\chi$, which may (for example) correspond to the dimensions of indices within the tensor network. Assuming that the coefficients of this cost polynomial are small, costs which scale as excessively large powers of $\chi$ may then reflect an algorithm which pushes the limits of computational feasibility.

The trade-off between sophistication of an Ansatz and the associated cost of the update algorithm represents a critical tension in the development of novel tensor network algorithms. For example, the structure of the 4:1 2D MERA \cite{cincio2008,evenbly2009} indicates that this particular Ansatz will provide a powerful representation of highly-entangled 2D systems, and this has been confirmed analytically in \rcite{aguado2008} where it is shown to furnish a compact and physically meaningful description of the toric code. However, numerical computations using this Ansatz are hindered by an update cost of $\OO{\chi^{26}}$. 
The ability to quickly and conveniently determine the cost of contracting different tensor networks is therefore of great importance to researchers employed in the development of novel tensor network algorithms.

\begin{figure}
\includegraphics[width=246.0pt]{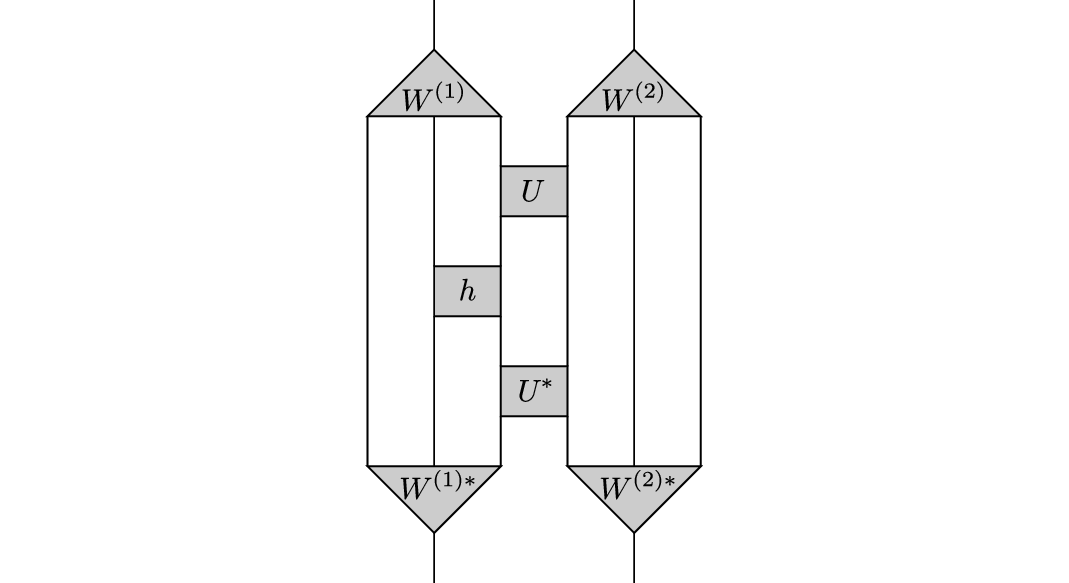}
\caption{\label{fig:31MERA}Graphical representation of one of the tensor networks which must be contracted during variational optimisation of the 3:1 MERA \pcite{evenbly2009,vidal2010}. The individual tensors are labelled for subsequent reference, and the tensor expression corresponding to this diagram is given in \pEref{eq:31MERA} of \psref{sec:results}. %
}
\end{figure}

Even when the cost of updating a tensor network algorithm is known, the implementation of these algorithms is frequently a non-trivial affair. The process of contracting a tensor network may always be optimally realised as a series of pairwise contractions (see \aref{sec:pairwise}), and the overall cost of contracting the network is highly dependent upon the sequence in which these contractions are carried out. For instance, the network shown in \fref{fig:31MERA} is one network which must be contracted during the variational optimisation of the 3:1 1D MERA \cite{evenbly2009,vidal2010}. 
If each index has dimension $\chi$ then the most efficient contraction sequences yield a cost of $\OO{\chi^8}$, for example
\begin{equation}
(((((hU)U^*)W^{(1)})W^{(1)*})(W^{(2)}W^{(2)*})),
\end{equation}
but careless choices of sequence can yield costs as high as $\OO{\chi^{11}}$, for example
\begin{equation}
(((W^{(2)}U)h)(((W^{(1)}W^{(1)*})U^*)W^{(2)*})).
\end{equation} 
Thus the ability to determine the optimal contraction sequence for a tensor network is as important to those implementing pre-existing tensor network algorithms as it is to those developing them, if the algorithms are to be implemented in a computationally efficient manner.

Until recently, manual optimisation has largely been the preferred approach within the tensor network community. However, the increasing complexity of tensor network Ans\"atze (e.g.~2D MERA \cite{cincio2008,evenbly2009,evenbly2009b}, %
branching MERA \cite{evenbly2011}) renders this approach increasingly time-consuming and, with an exhaustive search out of the question, it may be difficult to be certain that an optimal contraction sequence has indeed been identified. This situation is exemplified by the 4:1 2D MERA of Refs.~\onlinecite{cincio2008,evenbly2009}, which was believed for a number of years to have a minimal contraction cost of $\OO{\chi^{28}}$ operations and has only recently been shown to be contractible for only $\OO{\chi^{26}}$ \cite{pfeifer2013}. As tensor network algorithms increase in size and complexity, and implementations begin the transition from single-node to parallel computing architectures, the limitations of manual optimisation are likely to become increasingly apparent.

\subsection{Floating point operations and tensor contractions\label{sec:flops}}

When working with tensor network algorithms, the primary objects of interest are single tensor networks made up of a large number of constituent tensors. The principal operation applied to these networks is the contraction of a pair of tensors to form a single tensor, e.g.
\begin{equation}
C_{ijk} = A_{ik}^l B_{lj}.
\end{equation}
Evaluation of $C_{ijk}$ proceeds by iteration over indices $i$, $j$, and $k$, and summation over index $l$. If the array representing tensor $C_{ijk}$ is initialised to zero, then construction of $C_{ijk}$ involves $\xi_i\xi_j\xi_k\xi_l$ floating-point multiplications and the same number of additions. With the number of multiplications and additions being equal, it is customary to count only multiplications, and the above calculation is described as having a cost of \emph{$\xi_i\xi_j\xi_k\xi_l$ operations}.

This may be contrasted with the operation counting adopted by the quantum chemistry and computer science communities. In quantum chemistry Ans\"atze, both tensor contraction and tensor summation play important roles and thus independent tallies must be kept of multiplication and addition operations. The calculation of $C_{ijk}$ given above may in principle be decomposed into two steps \cite{lam1997},
\begin{align}
\left(C^*\right)^l_{ijkl} &= A_{ik}^l\times B_{lj}\qquad\textrm{(no sum over $l$)}\label{eq:lunsummed}\\
C_{ijk} &= \sum_l \left(C^*\right)^l_{ijkl}
\end{align}
where the first calculation is iterated over $i$, $j$, $k$, and $l$ for $\xi_i\xi_j\xi_k\xi_l$ floating-point multiplication operations, and the second is iterated over $i$, $j$, and $k$, and for each set of values $\{i,j,k\}$ the right-hand side is summed over $l$ for a total of $\xi_i\xi_j\xi_k\xi_l$ floating-point addition operations. In practice there is seldom reason to generate the intermediate object $C^*$ explicitly.

It is noted in \rcite{lam1997} that following multiplication, it is never suboptimal to immediately sum over all repeated indices on the resulting object (for example, $l$ on $C^*$ in the above example). If this is performed, then the process of multiplication followed by summation over all repeated indices is directly equivalent to contraction of two tensors over all shared indices. 

In the present context of optimising the contraction of a single tensor network, all operations may therefore be understood as pairwise tensor contractions. It is consequently unnecessary to separately count floating-point additions and multiplications (as these numbers are always equal), so for simplicity we shall adopt the approach of the tensor network algorithm community by counting operations which consist of one floating-point multiplication and one floating-point addition apiece. Revisiting the simple examples of \fref{fig:examplenetworks}, evaluation of $C$ in \Eref{eq:AtimesB} therefore incurs a cost of $\xi_i\xi_j\xi_k$ operations, while evaluation of $c$ in \Eref{eq:vMw} attracts a cost dependent on the order of pairwise contractions. If one
begins by contracting $v$ with $M$,
\begin{equation}
x^j=v_iM^{ij},\qquad c=x^jw_j,
\end{equation}
then the cost of computing $c$ is $\xi_i\xi_j+\xi_j$ operations, whereas if one first contracts
$w$ with $M$,
\begin{equation}
y=M^{ij}w_j,\qquad c=v_iy^i,
\end{equation}
the cost is $\xi_i\xi_j+\xi_i$ operations.
Finally, if one begins by performing an outer product between $v$ and $w$,
\begin{equation}
z_{ij}=v_iw_j,\qquad c=z_{ij}M^{ij}
\end{equation}
then this approach incurs a cost of $2\xi_i\xi_j$.

\section{Single term optimisation\label{sec:singleterm}}

This Section discusses algorithms for the identification of an optimal (operation-minimising) contraction sequence for a single-term tensor expression. In describing these algorithms it is assumed that the expression contains no indices of dimension one, and that it may not be factorised into two disjoint parts sharing no common indices.

Regarding the first of these requirements, note that an index of dimension one may freely be deleted or reinserted at any time without changing the calculation described, and thus this condition may be imposed without loss of generality. 

Regarding the second, note that
for any network $\mc{N}$ which is factorisable into two or more disjoint subnetworks $\mc{N}_1,\mc{N}_2,\ldots$, e.g.
\begin{equation}
A^{ij}B^{l}C_{jk}D_{lm} = \left(A^{ij}C_{jk}\right)\left(B^l D_{lm}\right),\label{eq:disjoint}
\end{equation} 
and where contraction of these subnetworks yields tensors of dimension greater than one,
an optimal contraction sequence may always be constructed by sequentially concatenating optimal contraction sequences for each subnetwork. In these circumstances we lose no capability (and make substantial performance gains) by addressing each subnetwork individually.\footnote{\label{fn:scalar}Where one or more subnetworks does contract to a tensor of dimension one (i.e.,~a single number, a scalar), the situation is a little more complex. In theory the optimal time to multiply by this number may arise part-way through the evaluation of another tensor network, if the contraction of that network yields an especially small intermediate object. In practice the best time to multiply by this number is usually during the calculation of human-readable results, which are seldom expressed in terms of large, multi-dimensional tensors. The problem of determining the optimal time to multiply by a scalar is therefore frequently outside the context addressed by the present paper, but frequently admits the same approach as discussed above, namely independent contraction of each individual subnetwork.}
Note also that the disjoint or non-disjoint nature of the network is to be assessed only after any indices of dimension one have been deleted.

Finally, it is assumed that no index appears more than once on any of the initial tensors. As mentioned in \rcite{lam1997}, it is always optimal to perform the sums over such indices immediately, and thus it is assumed that these have already been performed before the algorithms described in this Section are invoked.

\subsection{Existing approaches\label{sec:existing}}

\subsubsection{Depth-first constructive approach\label{sec:depthalg}}

We begin with a review of the approach to single-term optimisation first described in \rcite{lam1997}. 
Slightly paraphrased, the algorithm proceeds as follows:
\begin{enumerate}
\item Compute a single contraction sequence as follows: \label{step:singleseq}
\begin{enumerate}
\item
\begin{enumerate}
\item Let $T_1,\ldots,T_n$ be a list of $n$ tensors.
\item Let $d=n$ be a counter enumerating the total number of unique objects.
\item Let $c$ be a counter initialised to zero.
\end{enumerate}
\item \label{step:repeat}
\begin{enumerate}
\item Increment $d$. 
\item Choose a pair of tensors $\{T_a,T_b\}$ such that the set indices appearing on $T_a$ is identical to that appearing on $T_b$, or if no such pair exists, choose a pair such that $a<b$ and $b>c$.
\item Contract these two tensors over all common indices (if any). Call the resulting object $T_d$.
\item Remove tensors $T_a$ and $T_b$ from the list of tensors.
\item Append $T_d$ to the list of tensors.
\item Set $c$ to $a$.
\end{enumerate}
\item Repeat step~\ref{step:repeat} until no further actions can be applied. Determine the cost of the contraction sequence performed.
\end{enumerate}
\item Repeat step~\ref{step:singleseq} while iterating through all possible choices of pairs $\{T_a,T_b\}$ using a depth-first approach, to explore the space of all possible contraction sequences without duplication. Note the cheapest contraction sequence thus found.
\end{enumerate}

Note, in particular, that tensors $T_a$ and $T_b$ need not share any common indices at all. A contraction where tensors do not share any common indices is termed an \emph{outer product}, and there exist tensor networks for which an outer product is a necessary part of the optimal contraction sequence. A simple example is given by
\begin{equation}
D_k = A^iB^jC_{ijk}
\end{equation}
for $\xi_k>\xi_j$ and $\xi_k>\xi_i$. Writing $\fuse{X}{Y}$ to denote the contraction of a tensor $X$ with a tensor $Y$, the sequence $\fuse{\fuse{A}{B}}{C}$ is seen to be cheaper than either $\fuse{\fuse{A}{C}}{B}$ or $\fuse{\fuse{B}{C}}{A}$.

\subsubsection{Breadth-first constructive approach\label{sec:breadthalg}}

A slightly more sophisticated approach may be found in \rcite{hartono2005}, and has been incorporated into the Tensor Contraction Engine \cite{hirata2003}, a tool for optimising the evaluation of tensor expressions which enjoys widespread use in the quantum chemistry community. Whereas implementation of the depth-first search of \sref{sec:depthalg} will typically be achieved through the use of recursion, the breadth-first approach described in \rcite{hartono2005} is %
iterative and may be summarised as follows:

\begin{enumerate}
\item Let $S_1=\{T_1,\ldots,T_n\}$ be the set of $n$ tensors which make up network $\mc{N}$.
\item Let $c$ be a counter running from 2 to $n$. For each value of $c$:
\begin{enumerate}
\item Let $S_c$ be the set of all objects made up by contracting together $c$ unique tensors from $S_1$.
\item For each pair of sets $S_d$, $S_{c-d}$, $1\leq d\leq \lfloor \frac{c}{2}\rfloor$, and for each $T_a\in S_d$, $T_b\in S_{c-d}$ such that each element of $S_1$ appears at most once in $\fuse{T_a}{T_b}$: 
\begin{enumerate}
\item Determine the cost $\mu$ of contracting $T_a$ with $T_b$.
\item Where $T_a$ and/or $T_b$ do not belong to $S_1$, add to $\mu$ the previously-determined cost of constructing $T_a$ and/or $T_b$ as appropriate.
\item Let the contraction sequence $\mc{Q}$ for constructing this object be written $\mc{Q}=\fuse{T_a}{T_b}$. Where $T_a$ and/or $T_b$ do not belong to $S_1$, optimal contraction sequences for $T_a$ and $T_b$ will have been previously recorded. In $\mc{Q}$, replace each appearance of $T_a$ and/or $T_b$ with these optimal contraction sequences.
\item Locate the object in $S_c$ which corresponds to $\fuse{T_a}{T_b}$. If $\mu$ is the cheapest known cost for constructing this object, record the cost $\mu$ and the associated contraction sequence $\mc{Q}$ against this object.
\end{enumerate}
\end{enumerate}
\item The optimal cost $\mu_\mrm{best}$ and a sequence $\mc{Q}_\mrm{best}$ which realises this are recorded against the only element in $S_n$.
\end{enumerate}

\subsubsection{Dynamic programming\label{sec:DPalg}}

Finally, as with the matrix-chain problem discussed in \aref{sec:matmult}, an exhaustive search of all possible contraction sequences may also be performed using dynamic programming with memoisation: 
\begin{enumerate}
\item Is network $\mc{N}$ trivial (one tensor)? If so, let $T$ denote this tensor, and return zero cost and a contraction sequence $\mc{Q}=T$.
\item Has network $\mc{N}$ been costed before? If so, return the known optimal cost and sequence, previously recorded as $\mu_\mrm{best}(\mc{N})$ and $\mc{Q}_\mrm{best}(\mc{N})$.
\item Otherwise, for each bipartition of $\mc{N}$ into two subnetworks $\mc{N}_1$ and $\mc{N}_2$ (either or both of which may be disjoint):
\begin{enumerate}
\item Invoke this algorithm twice more, to obtain the optimal cost and sequence for each of $\mc{N}_1$ and $\mc{N}_2$.
\item Let $T_1$ be the tensor obtained on fully contracting $\mc{N}_1$. Let $T_2$ be the tensor obtained on fully contracting $\mc{N}_2$. Let $\mu$ be the cost of the contraction $\fuse{T_1}{T_2}$.
\item Add to $\mu$ the optimal cost for contracting $\mc{N}_1$ and the optimal cost for contracting $\mc{N}_2$.
\item Define a sequence $\mc{Q}=(\mc{Q}_1\mc{Q}_2)$ where $\mc{Q}_a$ is the optimal sequence for contracting subnetwork $\mc{N}_a$.
\item If $\mu$ is the best cost observed, record the values of $\mu$ and $\mc{Q}$ as $\mu_\mrm{best}(\mc{N})$ and $\mc{Q}_\mrm{best}(\mc{N})$, overwriting any previous values thus recorded.
\end{enumerate}
\item Return $\mu_\mrm{best}(\mc{N})$ and $\mc{Q}_\mrm{best}(\mc{N})$.
\end{enumerate}
Note the role of memoisation in preventing unnecessary duplication of efforts: While iterating over all possible bifurications of networks at all levels of recursion, the same subnetworks will frequently be encountered on many different occasions. As a simple example, given a network $\mc{N}=\{ABCDEF\}$ the subnetwork $\{BC\}$ may be encountered by splitting $\mc{N}$ into $\{ADEF\}$ and $\{BC\}$, or by splitting $\mc{N}$ into $\{ABC\}$ and $\{DEF\}$ then splitting $\{ABC\}$ into $A$ and $\{BC\}$, or in numerous other ways.

\subsection{Beyond existing approaches\label{sec:beyond}}

While the approaches described in \sref{sec:existing} are adequate for small tensor networks, they become rapidly more expensive as $n$, the number of tensors in the network, increases. This Section introduces two modifications to the search process. The first, described in \sref{sec:costcap}, represents a re-ordering of the search process in favour of cheapest-first. As the search may now terminate once a contraction sequence has been identified, this approach prunes the search tree to exclude all contraction sequences costing more than the optimal contraction cost of the tensor network. The second modification, described in \sref{sec:restrictops}, excludes large numbers of outer product contractions which are shown to be unnecessary for the identification of an optimal contraction sequence.

\subsubsection{Cost capping\label{sec:costcap}}

The refinement described in this Section may be applied to any of the search algorithms described in \sref{sec:existing}. However, to obtain the most benefit it should be applied to an algorithm incorporating some form of memoisation, either explicitly (the dynamic programming algorithm of \sref{sec:DPalg}) or implicitly (the recording of objects and their associated costs and sequences in the breadth-first algorithm of \sref{sec:breadthalg}). Of the two, the breadth-first algorithm of \sref{sec:breadthalg} is found to perform significantly better with the cost capping refinement as the initial space to be explored [$n(n-1)/2$  possible pairwise tensor contractions] is both smaller than, and typically contains less expensive (and therefore more relevant) elements than the initial space of the dynamic programming algorithm ($2^{n-1}$ bifurcations of the initial network $\mc{N}$).\footnote{Example results for a dynamic programming algorithm with cost capping and some restrictions on outer products may be found in Table~V of \prcite{pfeifer2013}. They are most appropriately compared with column ``\protect{$\mu_\mrm{cap}$}'' of \ptref{tab:results} in \psref{sec:results}, with performance of the breadth-first code being between eight and sixty times faster.}

To implement cost capping in the breadth-first algorithm of \sref{sec:breadthalg}, two modifications are required. First, the sets $S_c$ are initialised empty and are only populated as contraction sequences which yield their elements are identified. Second, pairwise contractions are rejected if their cost exceeds some maximal value $\mu_\mrm{cap}$. By choosing a value of $\mu_\mrm{cap}$ which is sufficiently low, the first invocation of the modified breadth-first search can be guaranteed to terminate with $S_n$ empty, i.e.,~without constructing an object containing all of the tensors in $S_1$. If the value of $\mu_\mrm{cap}$ is then increased, and the former value is stored in $\mu_\mrm{old}$, the existing contents of the sets $S_i$ act as memos for the second invocation of the breadth-first search. On this second invocation it is only necessary to consider contractions where the cost $\mu$ of $(T_aT_b)$ satisfies $\mu_\mrm{old}<\mu\leq\mu_\mrm{cap}$, or where $\mu\leq\mu_\mrm{old}$ and one or both of $T_a$ and $T_b$ was itself only constructed during the second invocation. If, following the second invocation, $S_n$ is still empty, $\mu_\mrm{cap}$ is increased once more, $\mu_\mrm{old}$ is updated, and the process is repeated. The net result is to yield a cheapest-first construction process where no object is constructed which costs significantly more than the contraction of network $\mc{N}$.

For tensor network algorithms, it is common to define bond dimensions in terms of some refinement parameter $\chi$, assumed to be large. In this context, $\mu_\mrm{cap}$ may be taken as a bound on the maximum power of $\chi$ which appears in the algebraic expression for the contraction cost. Alternatively one may supply numeric values for the index dimensions and $\mu_\mrm{cap}$ then comprises a cap on the actual contraction cost.

Let $\xi_\mrm{min}$ denote the dimension of the smallest index in the network, let $\mu_\mrm{cap}^{(i)}$ denote the value of $\mu_\mrm{cap}$ on invocation $i$ of the algorithm, and let $\mu_\mrm{reject}^{(i)}$ represent the cost of the cheapest rejected pairwise contraction on iteration $i-1$.
We obtained good performance by initialising $\mu^{(1)}_\mrm{cap}$ to 1 and requiring that each subsequent $\mu^{(i)}_\mrm{cap}=\xi_\mrm{min}\mu_\mrm{cap}^{(i-1)}$, except when $\mu_\mrm{reject}^{(i)}>\xi_\mrm{min}\mu_\mrm{cap}^{(i-1)}$, for which we assign $\mu_\mrm{cap}^{(i)} = \mu_\mrm{reject}^{(i-1)}$. However, for networks with many indices of some dimension $\xi$ and a much smaller number of indices of dimension $\xi_\mrm{min}\ll\xi$ we acknowledge that it may be preferable to instigate a more rapid increase in $\xi_\mrm{min}$.

Writing $\ttt{cost}[\fuse{A}{B}]$ for the cost of contracting tensor $A$ with tensor $B$, a cost-capped breadth-first algorithm may be realised as follows:

\begin{enumerate}
\item Let $S_1=\{T_1,\ldots,T_n\}$ be the set of $n$ tensors which make up network $\mc{N}$.
\item Flag each tensor in $S_1$ as ``old''.
\item Let $\{S_i~|~i\in\mbb{Z},~2\leq i\leq n\}$, be empty sets.
\item Let $\mu_\mrm{cap}=1$, let $\mu_\mrm{old}=0$, and let $\xi_\mrm{min}$ be the dimension of the smallest index.
\item While $S_n$ is empty:
\begin{enumerate}
\item Let $\mu_\mrm{next}=\infty$.
\item Let $c$ be a counter running from 2 to $n$.\\For each value of $c$,\\and each pair of sets $S_d$, $S_{c-d}$, $1\leq d\leq \lfloor \frac{c}{2}\rfloor$,\\and each $T_a\in S_d$, $T_b\in S_{c-d}$ such that each element of $S_1$ appears at most once in $\fuse{T_a}{T_b}$:
\begin{enumerate}
\item Let $\mu=\ttt{cost}[\fuse{T_a}{T_b}]$.
\item Where $T_a$ and/or $T_b$ do not belong to $S_1$, add to $\mu$ the previously-determined cost of constructing $T_a$ and/or $T_b$ as appropriate.
\item If either $T_a$ or $T_b$ is flagged as ``new'', let $\mu_0=0$. Otherwise, let $\mu_0=\mu_\mrm{old}$.
\item If $\mu>\mu_\mrm{cap}$ and $\mu<\mu_\mrm{next}$, let $\mu_\mrm{next}=\mu$.
\item If $\mu_0<\mu\leq\mu_\mrm{cap}$: 
\begin{enumerate}
\item Let the contraction sequence $\mc{Q}$ for constructing this object be written $\mc{Q}=\fuse{T_a}{T_b}$. Where $T_a$ and/or $T_b$ do not belong to $S_1$, the best-known contraction sequences for $T_a$ and $T_b$ will have been previously recorded. In $\mc{Q}$, replace each appearance of $T_a$ and/or $T_b$ with these optimal contraction sequences.
\item If no object corresponding to $\fuse{T_a}{T_b}$ has yet been created in $S_c$, create it. Otherwise, locate the object in $S_c$ which corresponds to $\fuse{T_a}{T_b}$.
\item If $\mu$ is the cheapest known cost for constructing this object then record the cost $\mu$ and the associated contraction sequence $\mc{Q}$ against this object, and flag the object as ``new''.
\end{enumerate}
\end{enumerate}
\item Let $\mu_\mrm{old}=\mu_\mrm{cap}$.
\item Set $\mu_\mrm{cap}$ equal to the larger of $\mu_\mrm{next}$ and $\xi_\mrm{min}\mu_\mrm{cap}$.
\item Flag all tensors in all $S_i$ as ``old''.
\end{enumerate}
\item The optimal cost $\mu_\mrm{best}$ and a sequence $\mc{Q}_\mrm{best}$ which realises this are recorded against the only element in $S_n$.
\end{enumerate}

\subsubsection{Restricting outer products\label{sec:restrictops}}

\paragraph{Overview:\label{sec:roverview}}
As noted in \sref{sec:depthalg}, there exist tensor networks for which outer products form a necessary part of the optimal contraction sequence. Consequently, when iterating over pairs of tensors to contract in the breadth-first search algorithm, this iteration must include not only pairs of tensors which share one or more common indices, but also pairs of tensors which do not. 
In Secs.~\ref{sec:tensorsinOP}--\ref{sec:tensorswithOP} a number of criteria are introduced for the rejection of outer product operations, and it is proven that application of these criteria will never prevent the identification of an optimal contraction sequence. Use is made of the following two lemmas: %

\begin{lemma}[Combination of shared indices]
Suppose there exist two tensors, $P$ and $Q$, which share two or more indices. Let us denote these indices $\{a_1,a_2,\ldots,a_n\}$. We may combine these indices into a single index $a$. \emph{Proof:} Let $a$ enumerate all possible sets of values %
$\{a_1,a_2,\ldots,a_n\}$.\hfill$\square$\label{lemma:sharedindices}
\end{lemma}
\begin{lemma}[Combination of external indices]
Next, suppose we are only interested in the subnet comprising tensors $P$ and $Q$. Let us denote all indices on $P$ which do not connect to $Q$ by $\{b_1,b_2,\ldots,b_n\}$. We may similarly replace all of these indices by a single combined index $b$. \emph{Proof:} Let $b$ %
enumerate all possible sets of values $\{b_1,b_2,\ldots,b_n\}$.\hfill$\square$\label{lemma:extindices}
\end{lemma}

\paragraph{Constraints on tensors participating in the outer product:\label{sec:tensorsinOP}}
Let $A$ and $B$ be two tensors which we wish to contract together as part of our contraction sequence, and which share no common indices. For a non-disjoint network $\mc{N}$, 
object $\fuse{A}{B}$ will necessarily be subsequently contracted with some other tensor $C$. If $C$ is not a fundamental tensor (i.e.~one belonging to $S_1$), we may always implement our contraction sequence such that $C$ is assembled before contraction $\fuse{A}{B}$ is performed. We will denote the subnetwork comprising tensors $A$, $B$, and $C$ by $\mc{N}_{ABC}$.

\begin{figure}
\includegraphics[width=246.0pt]{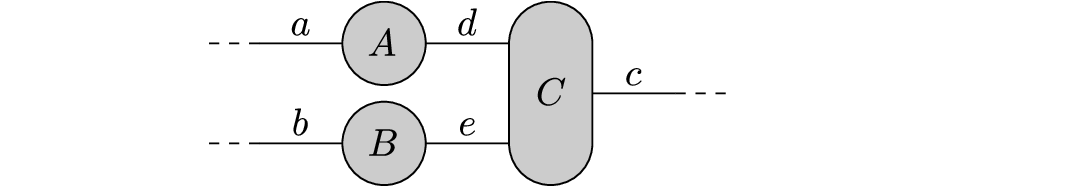}
\caption{Two tensors $A$ and $B$ share no common index; should they be combined using an outer product before contracting with tensor~$C$? Letters $a,\ldots,e$ denote indices of dimensions $\xa,\ldots,\xe$ respectively. In \psref{sec:tensorsinOP} it is shown that when searching for an optimal sequence we need only consider contraction $(AB)$ if indices $a,\ldots,e$ satisfy the conditions given in \pfref{fig:ABop1}.\label{fig:ABop}}
\end{figure}

By combining indices in the manner described in Lemmas~\ref{lemma:sharedindices}--\ref{lemma:extindices}, %
the most general form of subnetwork $\mc{N}_{ABC}$ is that given in \fref{fig:ABop}. 
\begin{figure}
\includegraphics[width=246.0pt]{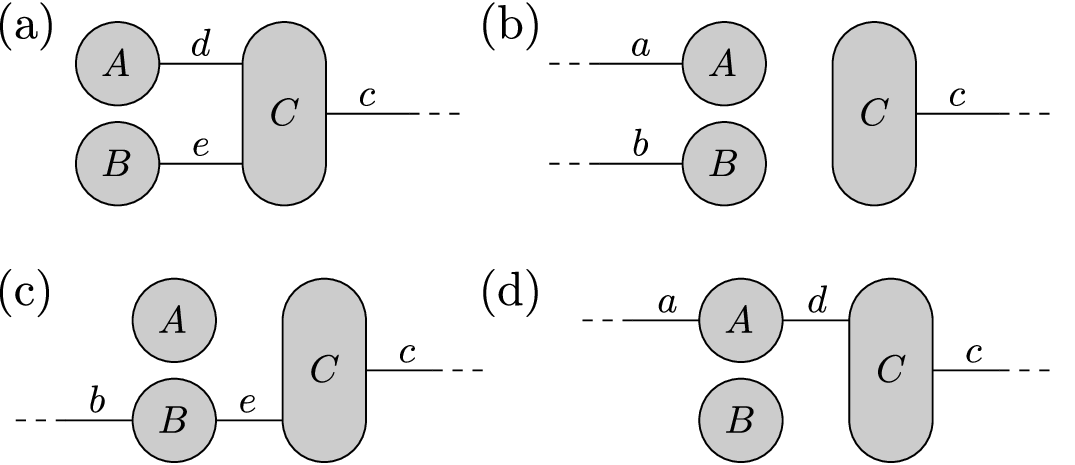}
\caption{Scenarios for which the outer product $((AB)C)$ may be of lower cost than either $((AC)B)$ or $((BC)A)$: (a)~$\xa=\xb=1$. (b)~$\xd=\xe=1$. (c)~$\xa=\xd=1$. (d)~$\xb=\xe=1$.\label{fig:ABop1}}
\end{figure}%
If an outer product between $A$ and $B$ is a necessary part of the contraction sequence it follows that sequence $((AB)C)$ is cheaper than either $((AC)B)$ or $((BC)A)$. Comparison with sequence $((AC)B)$ gives us the inequality
\begin{equation}
\begin{split}
\ttt{cost}[((AB)C)]&<\ttt{cost}[((AC)B)]\\
\xa\xb\xd\xe+\xa\xb\xc\xd\xe&<\xa\xc\xd\xe+\xa\xb\xc\xe
\end{split}\label{eq:ineq1}
\end{equation}
simplifying to
\begin{equation}
\xb\xd<\xc(\xb+\xd-\xb\xd).\label{eq:ABopconstraint1}
\end{equation}
Since all dimensions are positive integers, this requires 
\begin{equation}
\xb+\xd-\xb\xd>0
\end{equation}
and may only be satisfied if $\xb=1$ or $\xd=1$. The condition obtained from sequence $((BC)A)$ is equivalent under the exchanges 
\begin{equation}
A\Leftrightarrow B,~a\Leftrightarrow b,~d\Leftrightarrow e, 
\end{equation}
so yields 
\begin{equation}
\xa\xe<\xc(\xa+\xe-\xa\xe)\label{eq:ABopconstraint2}
\end{equation}
which requires either $\xa=1$ or $\xe=1$. There are consequently only three scenarios in which sequence $((AB)C)$ is superior to either $((AC)B)$ or $((BC)A)$, and thus an outer product may be required:
\begin{enumerate}
\item $\xa=\xb=1$: Following an outer product between $A$ and $B$, all indices on the resulting tensor $(AB)$ are shared with a single tensor $C$ [\fref{fig:ABop1}(a)].\label{item:(AB)noopC}
\item $\xd=\xe=1$: Following an outer product between $A$ and $B$, the next operation in the contraction sequence is a further outer product between $(AB)$ and $C$ [\fref{fig:ABop1}(b)]. 
\label{item:(AB)opC}
\item $\xa=\xd=1$ or $\xb=\xe=1$: Either $A$ or $B$ is a scalar [\fref{fig:ABop1}(c)-(d)]. Provided network $\mc{N}$ is non-disjoint, these scenarios will never occur and so may be disregarded.\label{item:scenario3}
\end{enumerate}

Regarding \scref{item:(AB)noopC}, substitution of $\xa=\xb=1$ into \Erefs{eq:ABopconstraint1}{eq:ABopconstraint2} reveals the further constraints 
\begin{equation}
\xc>\xd\qquad\xc>\xe.\label{eq:OPconds} 
\end{equation}
As \scref{item:(AB)noopC} is the only admissible scenario in which the outer product object $(AB)$ shares common indices with tensor $C$, we may conclude that contraction of an outer product object $(AB)$ with a tensor $C$ is only necessary if it satisfies the following constraints:
\begin{enumerate}
\item All indices on tensor $(AB)$ must be shared with tensor $C$.
\item The indices on $(AB)$ and $C$ must satisfy \Eref{eq:OPconds}.
\end{enumerate}
Note in particular that when $\xc=\xd$ or $\xc=\xe$, the cost of sequence $((AB)C)$ is equal to that of either $((AC)B)$ or $((BC)A)$ respectively, but the latter two sequences do not involve outer products. By discarding sequence $((AB)C)$ when these are equal while retaining either or both of $((AC)B)$ and $((BC)A)$ we guarantee contraction of $\mc{N}_{ABC}$ at a cost equal to or less than $\ttt{cost}[((AB)C)]$, without requiring an outer product.

Regarding \scref{item:(AB)opC}, note that in this context all three sequences $((AB)C)$, $((AC)B)$, and $((BC)A)$ involve outer products. Substitution of $\xd=\xe=1$ into \Erefs{eq:ABopconstraint1}{eq:ABopconstraint2} yields
\begin{equation}
\xc>\xa\qquad\xc>\xb\label{eq:OPconds2}
\end{equation}
as the conditions under which sequence $((AB)C)$ is superior to sequence $((AC)B)$ and $((BC)A)$. However, if $\ttt{cost}[((AB)C)]=\ttt{cost}[((AC)B)]$ then enforcement of \Eref{eq:OPconds2} as a strict inequality will eliminate sequence $((AB)C)$, but $(AC)$ in sequence $((AC)B)$ is also an outer product and under permutation of labels
\begin{equation}
B\Leftrightarrow C,~b\Leftrightarrow c
\end{equation}
this constraint becomes
\begin{equation}
\xb>\xa\qquad\xb>\xc\label{eq:OPconds2a}
\end{equation}
and sequence $((AC)B)$ will also be rejected. This may be contrasted with \scref{item:(AB)noopC} where sequence $((AC)B)$ is not an outer product, is therefore not at risk of being rejected, and hence at least one sequence for contracting $\mc{N}_{ABC}$ at a cost equal or less than that of $((AB)C)$ is retained. In order to ensure that the optimal contraction sequence for subnetwork $\mc{N}_{ABC}$ is not inadvertently discarded when $\ttt{cost}[((AB)C)]$ is optimal and is equal to $\ttt{cost}[((AC)B)]$ or $\ttt{cost}[((BC)A)]$, the correct constraint to apply when contraction $((AB)C)$ is an outer product is 
\begin{equation}
\xc\geq\xa\qquad\xc\geq\xb,\label{eq:OPconds2*}
\end{equation}
corresponding to the requirement that multiple outer products be performed in non-descending order of tensor dimension.

\paragraph{Constraints on tensors contracting with an outer product:\label{sec:tensorswithOP}}
We now consider in further detail the situation where contraction of $(AB)$ with $C$ is not an outer product and tensor $C$ is composite. In \sref{sec:tensorsinOP} we assumed that a composite tensor~$C$ was always constructed before performing the outer product $(AB)$. 
If the final two tensors involved in the construction of $C$ are denoted $D$ and $E$, this corresponds to evaluating $(DE)$ before $(AB)$ in the sequence $((AB)(DE))$. However, with no impact on cost we might equally well choose to evaluate $(AB)$ before $(DE)$.

\begin{figure}
\includegraphics[width=246.0pt]{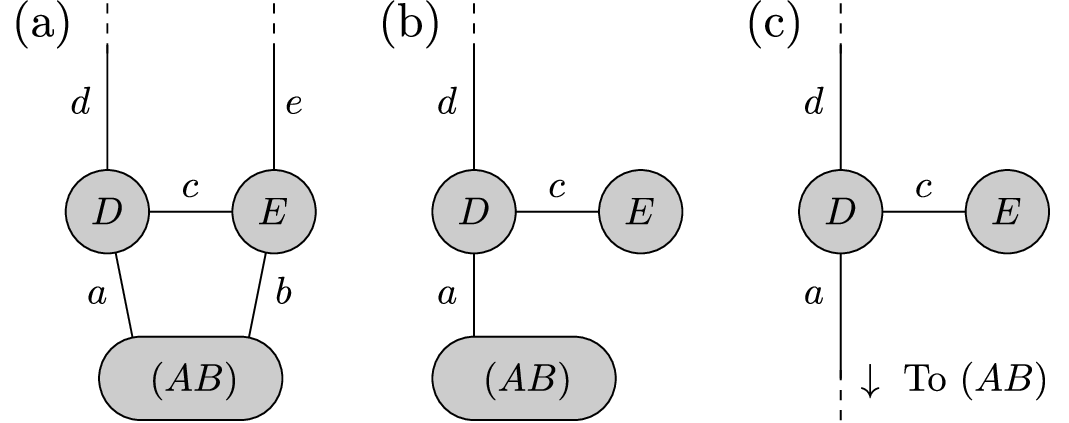}
\caption{(a)~Let tensor~$C$ in \pfref{fig:ABop1}(a) be composite, with the final step in its construction being the contraction together of two tensors $D$ and $E$. If contraction $(AB)$ is performed prior to the final step in the construction of $C$, the most general form of network $\mc{N}_{(AB)DE}$ is as shown. (b)~If contraction sequence $((AB)(DE))$ is superior to both $(((AB)D)E)$ and $(((AB)E)D)$, diagram~(a) must reduce to the form shown here (up to an exchange of labels $D\leftrightarrow E$, $a\rightarrow b$, $d\rightarrow e$). (c)~Consequently, in \pscref{item:(AB)noopC} of \psref{sec:tensorsinOP} either tensor~$C$ is not composite, or the final two tensors in the optimal construction of $C$ admit the form shown here. Note in particular that for a non-disjoint network, tensor $C$ in \pscref{item:(AB)noopC} is consequently never the result of an outer product.\label{fig:(AB)DE}}
\end{figure}

Consider now a situation involving composite $C$ where optimal contraction of $\mc{N}_{ABC}$ is only achieved through sequence $((AB)C)$, and where the optimal sequence for the construction of $C$ is unique.
If contraction of $(AB)$ with $(DE)$ is not an outer product, then by virtue of \scref{item:(AB)noopC} the most general form of a subnetwork comprising tensors $(AB)$, $D$, and $E$ is given by \fref{fig:(AB)DE}(a).

In seeking the optimal contraction sequence for the entire tensor network of which \fref{fig:(AB)DE}(a) is a part, we only need to consider $((AB)C)$ as a possible optimal sequence %
for subnetwork $\mc{N}_{ABC}$, and thus $((AB)(DE))$ as a possible optimal sequence for $\mc{N}_{ABDE}$, if sequence $((AB)(DE))$ is cheaper than both $(((AB)D)E)$ and $(((AB)E)D)$. %
Comparison of $((AB)(DE))$ with $(((AB)D)E)$ yields the inequality 
\begin{equation}
\begin{split}
\ttt{cost}[((AB)(DE))]&<\ttt{cost}[(((AB)D)E)]\\
\xa\xb\xc\xd\xe+\xa\xb\xd\xe&<\xa\xb\xc\xd+\xb\xc\xd\xe
\end{split}\label{eq:ABDEineq1}
\end{equation}
\begin{equation}
\Rightarrow \xa\xb\xd\xe<\xc(\xa\xb\xd+\xb\xd\xe-\xa\xb\xd\xe),
\end{equation}
and because all index dimensions are positive integers this requires
\begin{equation}
\begin{split}
\xa\xb\xd+\xb\xd\xe&>\xa\xb\xd\xe\\
\Rightarrow\quad\xa+\xe&>\xa\xe
\end{split}
\end{equation}
which can only be satisfied for $\xa=1$ or $\xe=1$. Similarly requiring 
\begin{equation}
\begin{split}
\ttt{cost}[((AB)(DE))]&<\ttt{cost}[(((AB)E)D)] \\
\xa\xb\xc\xd\xe+\xa\xb\xd\xe&<\xa\xb\xc\xe+\xa\xc\xd\xe
\end{split}\label{eq:ABDEineq2} 
\end{equation}
yields $\xb=1$ or $\xd=1$.

Letting $\xd=\xe=1$ yields a configuration for which the sequence $((AB)(DE))$ is prohibited by \scref{item:(AB)noopC} of \sref{sec:tensorsinOP}, and letting $\xa=\xb=1$ is prohibited by our assumption that contraction of $(AB)$ with $C$ is not an outer product. Consequently, sequence $((AB)(DE))$ is only preferable to both $(((AB)D)E)$ and $(((AB)E)D)$ if either $\xa=\xd=1$ or $\xb=\xe=1$, both corresponding (up to an interchange of labels) to the network given in \fref{fig:(AB)DE}(b). It therefore follows that, in addition to the constraints of \sref{sec:tensorsinOP}, we need only consider contractions $((AB)(DE))$ for $(AB)$ an outer product under the following circumstances:
\begin{enumerate}
\item When contraction of $(AB)$ with $(DE)$ is itself an outer product, or
\item when either tensor $D$ or tensor $E$ contributes no unsummed indices to tensor $(DE)$.
\end{enumerate}
Further, substituting $\xb=\xe=1$ back into \Erefs{eq:ABDEineq1}{eq:ABDEineq2} yields the inequalities
\begin{align}
\begin{split}
\xa\xc\xd+\xa\xd&<\xa\xc\xd+\xc\xd\\
\Rightarrow \xa&<\xc\label{eq:Cconstr2}
\end{split}\\
\begin{split}
\xa\xc\xd+\xa\xd&<\xa\xc+\xa\xc\xd\\
\Rightarrow \xd&<\xc\label{eq:Cconstr1}
\end{split}
\end{align}
respectively, which also imply
\begin{equation}
\xc^3>|D|.\label{eq:cubevsD}
\end{equation}
Given tensors $D$ and $E$ consistent with \fref{fig:(AB)DE}(c), the constraints of \Erefs{eq:Cconstr2}{eq:Cconstr1} are both necessary and sufficient to ensure the satisfaction of \Erefs{eq:ABDEineq1}{eq:ABDEineq2}, and thus it is only necessary to consider a composite tensor~$C$ in \scref{item:(AB)noopC} of \sref{sec:tensorsinOP} if these two inequalities are satisfied.

As a slight subtlety, note that contraction of $(AB)$ with $E$ is an outer product, and so for $\xd=\xc$ the sequence $(((AB)E)D)$ is \emph{a priori} excluded by \Eref{eq:OPconds}. It might therefore appear necessary to relax condition~\eref{eq:Cconstr1} to 
\begin{equation}
\xd\leq\xc, \label{eq:Cconstr3}
\end{equation}
permitting retention of sequence $((AB)(DE))$ as a non-outer-product alternative when $\xd=\xc$. In practice this situation never arises, as sequences involving the outer product $(AB)$ are never necessary for the optimal contraction of such a network. This result is demonstrated in~\aref{sec:tensorswithOP_noAB}.

We have now shown that if the optimal sequence for constructing tensor~$C$ is unique,
\fref{fig:(AB)DE}(c) and \Erefs{eq:Cconstr2}{eq:Cconstr1} define the situations under which it is necessary to consider a composite tensor~$C$ in \scref{item:(AB)noopC} of \sref{sec:tensorsinOP}. Now consider situations where there exist multiple contraction sequences of optimal cost for the construction of tensor~$C$. If \emph{any} of these sequences are inconsistent with \fref{fig:(AB)DE}(c) or \Erefs{eq:Cconstr2}{eq:Cconstr1} then there exists a contraction sequence for subnetwork $\mc{N}_{ABDE}$ which is cheaper than $((AB)(DE))$, and we may omit consideration of sequences for the full network which involve the contraction of tensor $(AB)$ with tensor $(DE)$. It is therefore only necessary to consider a composite tensor for the role of tensor~$C$ in \scref{item:(AB)noopC} in \sref{sec:tensorsinOP} if \emph{no} construction of optimal cost for tensor~$C$ exists which is inconsistent with \fref{fig:(AB)DE}(c), and the \emph{smallest} value of $\xc$ encountered in any of these sequences satisfies \Erefs{eq:Cconstr2}{eq:Cconstr1}.

Finally, we confirm that if the optimal sequence for tensor~$C$ is inconsistent with \fref{fig:(AB)DE}(c) or \Erefs{eq:Cconstr2}{eq:Cconstr1} but a suboptimal sequence is consistent with these constraints, we still need not consider sequence $((AB)C)$. This follows immediately from the observation that a suboptimal construction for tensor~$C$ may never yield the optimal cost for sequence $((AB)C)$.

\paragraph{Preferred implementation of these constraints:\label{sec:preferredimp}}
Reduction of search runtime is best achieved by pruning as many branches of the search tree as possible, as early as possible. To this end, the preferred implementation of these conditions is not only as a restriction on contractions involving tensors which are outer products, but also as a restriction on which outer products are performed in the first place. To this end, a contraction $(AB)$ which is an outer product is only performed if a tensor $C$ is known such that contraction of $(AB)$ with $C$ is not an outer product, and:
\begin{enumerate}
\item Subnetwork $\mc{N}_{ABC}$ takes the form of \fref{fig:ABop1}(a), with indices satisfying \Eref{eq:OPconds}. \label{cond1}
\item Either tensor $C$ is a fundamental tensor (i.e.,~belongs to $S_1$), or the final two tensors in all optimal-cost constructions of $C$ take the form of tensors $D$ and $E$ in \fref{fig:(AB)DE}(b) with indices satisfying \Erefs{eq:Cconstr2}{eq:cubevsD}.\label{cond2}
\end{enumerate}
Note that multiple sequential outer products are not excluded by these constraints, e.g.
\begin{equation}
(((A_1A_2)B)C) 
\end{equation}
where both contractions in $((A_1A_2)B)$ are outer products. For such sequences, if the existence of tensor~$C$ satisfies conditions~\ref{cond1} and~\ref{cond2} for the contraction of $(A_1A_2)$ with $B$, then it also necessarily satisfies these conditions for the contraction of $A_1$ with $A_2$, even though this contraction is not necessarily immediately followed by contraction of $(A_1A_2)$ with tensor~$C$.

On the other hand, notice that \Eref{eq:Cconstr1} is \emph{not} enforced at this time. This is because it is possible for tensor~$((A_1A_2)B)$ to satisfy this inequality while tensor~$(A_1A_2)$ does not. The contraction $(A_1A_2)$ should not be excluded, however, as it is a necessary precursor to the outer product of tensor~$(A_1A_2)$ with $B$, and the subsequent contraction $(((A_1A_2)B)C)$ is then permitted. Enforcement of
\Eref{eq:Cconstr1} 
is therefore only performed at the time of
actually contracting an outer product object with another tensor under \scref{item:(AB)noopC} of \sref{sec:tensorsinOP}.

It should be noted that the pre-emptive enforcement of conditions described here augments, not supplants, the enforcement of these constraints during contraction of an outer product object~$(AB)$ with a tensor~$C$. 
The pre-emptive application of some constraints prevents the construction of some unnecessary outer product objects, but the active application of \emph{all} constraints in \sref{sec:restrictops} during the contraction of an outer product object $(AB)$ with another tensor is also necessary to ensure that the outer products which \emph{are} constructed only participate in contractions which are necessary according to the constraints determined above.
Such contractions must therefore comply with scenario~\ref{item:(AB)noopC} or \ref{item:(AB)opC} of \sref{sec:tensorsinOP}, along with constraints \eref{eq:OPconds} or \eref{eq:OPconds2*} respectively. If \scref{item:(AB)noopC} applies, then tensor $C$ must also either belong to $S_1$ or have an optimal contraction sequence whose final tensors take the form of \fref{fig:(AB)DE}(c) of \sref{sec:tensorswithOP}, with indices satisfying \Erefs{eq:Cconstr2}{eq:Cconstr1}. 

A pseudocode implementation of all these constraints is rather lengthy, and thus is given in \aref{sec:pseudocode}. %

\section{Results\label{sec:results}}

This Section compares the performance of the basic breadth-first algorithm of \sref{sec:breadthalg} with (i)~a breadth-first search supplemented by cost-capping (\sref{sec:costcap}), (ii)~a breadth-first algorithm supplemented by the restrictions on admissible outer products described in \sref{sec:restrictops}, and (iii)~the full algorithm of \aref{sec:pseudocode}, which incorporates both cost-capping and outer product restrictions. %
Each of these algorithms was applied to seven tensor networks of varying size and complexity, chosen to be representative of the tensor network contractions commonly encountered in condensed matter physics, where they arise from the following scenarios:
\begin{figure}
\includegraphics[width=246.0pt]{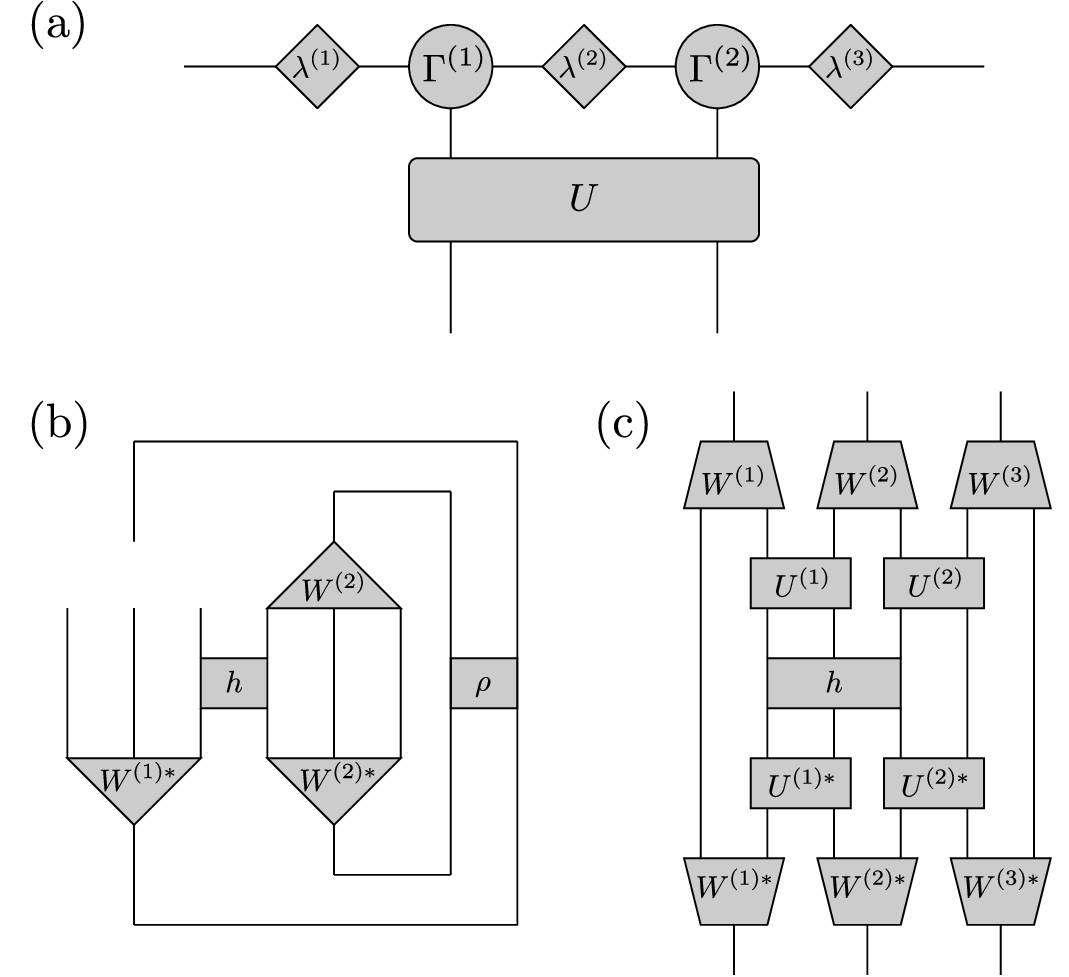}
\caption{Sample 1D tensor networks: 
(a)~Application of a time evolution gate in TEBD, corresponding to \pEref{eq:TEBD}. %
(b)~Calculation of the environment of a tensor in a 3:1 1D TTN, corresponding to \pEref{eq:1DTTN}. %
(c)~``Lifting'' a term in the Hamiltonian in a 2:1 1D MERA, corresponding to \pEref{eq:21MERA}. %
\label{fig:1DTNs}}
\end{figure}%
\begin{figure*}
\includegraphics[width=492.0pt]{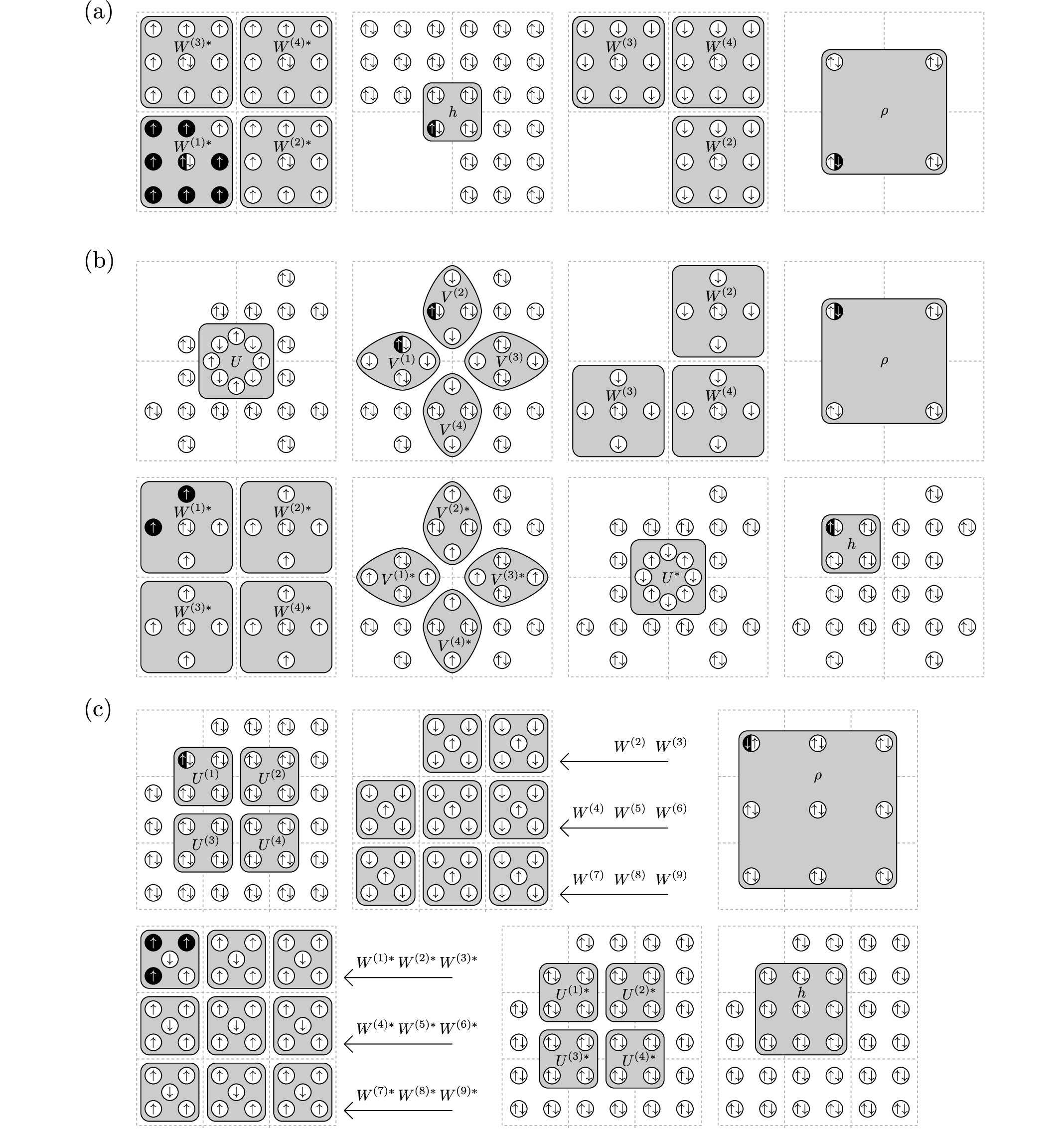}
\caption{Sample 2D tensor networks: 
(a)~Calculation of the environment of a tensor in a 9:1 2D TTN, corresponding to \pEref{eq:2DTTN}. %
(b)~Calculation of the environment of an isometry tensor in a 9:1 2D MERA, corresponding to \pEref{eq:2DTN1}. %
(c)~Calculation of the environment of an isometry tensor in a 4:1 2D MERA, corresponding to \pEref{eq:2DTN2}. %
For 2D tensor networks, a modified version of the diagrammatic notation is used in which the tensors of a diagram are arranged in layers. Unit cells on the lattice are marked on each layer as a reference guide, and circles containing a downward-pointing arrow represent indices passing to the layer below (drawn to the left), while those containing an upward-pointing arrow represent indices passing to the layer above (drawn to the right). Where a diagram extends over multiple rows, the layer at the left-hand end of one row is immediately above that at the right-hand end of the row below.
Circles containing both upward- and downward-pointing arrows indicate either a location on a tensor giving rise to both upward- and downward-going indices, or (if not located on a tensor) an index line passing through a given location in a layer. Paired indices are represented by black arrows on a white background, while unpaired indices are represented by a white arrow on black. and the diagrams are periodic, so that paired indices departing upwards from the top layer of a diagram enter the bottom layer from beneath, and vice versa.
\label{fig:2DTNs}}
\end{figure*}%
\begin{itemize}
\item \fref{fig:1DTNs}(a): Application of a time evolution gate in TEBD. Six tensors. The corresponding expression in index notation is
\begin{equation}
\lambda^{(1)}{}_{ab}\Gamma^{(1)}{}^{bd}_c\lambda^{(2)}{}_{de}\Gamma^{(2)}{}^{eg}_f\lambda^{(3)}{}_{gh}U^{cf}_{ij}\label{eq:TEBD}.
\end{equation}
\item \fref{fig:1DTNs}(b): Variational update of a tensor in a 3:1 1D TTN. Five tensors. The corresponding expression is
\begin{equation}
W^{(2)}{}^f_{cde}h^{kc}_{ab}W^{(1)*}{}^{ija}_hW^{(2)*}{}^{bde}_g\rho^{hg}_{lf}.\label{eq:1DTTN}
\end{equation}
\item \fref{fig:2DTNs}(a): Variational update of a tensor in a 9:1 2D TTN. Nine tensors. The corresponding expression is
\begin{equation}
\begin{split}
&W^{(2)}{}^c_{\varepsilon\zeta\eta\theta\iota\kappa j\lambda\mu}W^{(3)}{}^a_{ophqrstuv}W^{(4)}{}^b_{iwxyz\alpha\beta\gamma\delta} h^{\chi jhi}_{mnkl} \times\\
&W^{(1)*}{}^{\nu\xi\pi\rho\sigma\tau\upsilon\phi m}_f W^{(2)*}{}^{\varepsilon\zeta\eta\theta\iota\kappa n \lambda\mu}_g W^{(3)*}{}^{opkqrstuv}_d\times\\ 
&W^{(4)*}{}^{lwxyz\alpha\beta\gamma\delta}_e \rho^{fgde}_{\psi cab}.
\end{split}\label{eq:2DTTN}
\end{equation}
\item \fref{fig:31MERA}: ``Lifting'' the Hamiltonian in a 3:1 1D MERA. Seven tensors. The corresponding expression is
\begin{equation}
W^{(1)}{}^{a}_{efh}W^{(2)}{}^{b}_{lop}U^{hl}_{im}h^{fi}_{gj}U^*{}^{jm}_{kn}W^{(1)*}{}^{egk}_c W^{(2)*}{}^{nop}_d\label{eq:31MERA}.
\end{equation}
\item \fref{fig:1DTNs}(c): ``Lifting'' the Hamiltonian in a 2:1 1D MERA. Eleven tensors. The corresponding tensor expression is
\begin{equation}
\begin{split}
&W^{(1)}{}^u_{ae} W^{(2)}{}^v_{im} W^{(3)}{}^w_{pq} U^{(1)}{}^{ei}_{dh} U^{(2)}{}^{mp}_{lo} h^{dhl}_{cgk}\times\\
& U^{(1)*}{}^{cg}_{bf} U^{(2)*}{}^{ko}_{jn} W^{(1)*}{}^{ab}_r W^{(2)*}{}^{fj}_s W^{(3)*}{}^{nq}_t.
\end{split}\label{eq:21MERA}
\end{equation}
\item \fref{fig:2DTNs}(b): Variational update of an isometry tensor in a 9:1 2D MERA. Nineteen tensors. The corresponding tensor expression is
\begin{equation}
\begin{split}
&W^{(2)}{}^\alpha_{\delta\varepsilon\zeta\eta\theta}W^{(3)}{}^\beta_{\iota\kappa\lambda\mu\nu}W^{(4)}{}^\gamma_{\xi\pi\rho\sigma\tau}V^{(1)}{}^{\bar\pi\iota}_{\omega a b c}V^{(2)}{}^{\bar\xi\varepsilon}_{\upsilon\phi\chi\psi}\\
&\times V^{(3)}{}^{\theta\xi}_{d e f g}V^{(4)}{}^{\mu\pi}_{h i j k}U^{\psi b e h}_{l m n o}h^{\bar\nu\phi\omega l}_{p q r s} U^*{}^{s m n o}_{t u v w} V^{(1)*}{}^{r a u c}_{x y}   \\
&\times V^{(2)*}{}^{\upsilon q \chi t}_{z\bar\alpha} V^{(3)*}{}^{d v f g}_{\bar\beta\bar\gamma}V^{(4)*}{}^{w i j k}_{\bar\delta\bar\varepsilon}W^{(1)*}{}^{\bar\lambda\bar\mu p z x}_{\bar\zeta} \\
&\times W^{(2)*}{}^{\delta \bar\alpha \zeta\eta \bar\beta}_{\bar\eta} W^{(3)*}{}^{y \kappa\lambda\bar\delta\nu}_{\bar\theta} W^{(4)*}{}^{\bar\gamma\bar\varepsilon\rho\sigma\tau}_{\bar\iota} \rho^{\bar\zeta\bar\eta\bar\theta\bar\iota}_{\bar\kappa\alpha\beta\gamma}\label{eq:2DTN1}.
\end{split}
\end{equation}
\item \fref{fig:2DTNs}(c): Variational update of an isometry tensor in a 4:1 2D MERA. Twenty-seven tensors. The corresponding tensor expression is
\begin{equation}
\begin{split}
&W^{(2)}{}^{\bar\sigma}_{\bar{d}\bar{e}\xi\sigma}W^{(3)}{}^{\bar\tau}_{\bar f\bar g\tau\bar j}W^{(4)}{}^{\bar\upsilon}_{\bar k\pi\bar m\omega}W^{(5)}{}^{\bar\phi}_{\rho\upsilon a f}W^{(6)}{}^{\bar\chi}_{\phi\bar l g \bar n}\\
&\times W^{(7)}{}^{\bar\psi}_{\bar o b \bar q \bar r}W^{(8)}{}^{\bar\omega}_{c h \bar s\bar t}W^{(9)}{}^{\bar a}_{i \bar p\bar u\bar v} U^{(1)}{}^{\nu\xi\pi\rho}_{\bar\kappa\alpha\gamma\bar\lambda}U^{(2)}{}^{\sigma\tau\upsilon\phi}_{\beta\chi\bar\mu\psi}
\\
&\times U^{(3)}{}^{\omega abc}_{\bar\nu\bar\xi d e}U^{(4)}{}^{f g h i}_{\bar\pi j k l}h^{\bar\kappa\alpha\beta\gamma\bar\lambda\bar\mu\bar\nu\bar\xi\bar\pi}_{\delta\varepsilon\zeta\eta\theta\iota\kappa\lambda\mu}U^{(1)*}{}^{\delta\varepsilon\eta\theta}_{m n o p} U^{(2)*}{}^{\zeta\chi\iota\psi}_{s t u v}
\\
&\times U^{(3)*}{}^{\kappa\lambda de}_{yz\bar\alpha\bar\beta} U^{(4)*}{}^{\mu jkl}_{\bar\zeta\bar\eta\bar\theta\bar\iota}W^{(1)*}{}^{\bar b\bar c\bar h m}_{\bar w} W^{(2)*}{}^{\bar d\bar e ns}_{\bar x}  \\
&\times W^{(3)*}{}^{\bar f\bar g t\bar j}_qW^{(4)*}{}^{\bar k o \bar m y}_r W^{(5)*}{}^{puz\bar\zeta}_w W^{(6)*}{}^{v\bar l\bar\eta\bar n}_x  \\
&\times W^{(7)*}{}^{\bar o\bar\alpha\bar q\bar r}_{\bar\gamma}W^{(8)*}{}^{\bar\beta\bar\theta\bar s\bar t}_{\bar\delta}W^{(9)*}{}^{\bar\iota\bar p\bar u\bar v}_{\bar\varepsilon}\rho^{\bar w\bar x q r w x \bar\gamma\bar\delta\bar\varepsilon}_{\bar\rho\bar\sigma\bar\tau\bar\upsilon\bar\phi\bar\chi\bar\psi\bar\omega\bar a}
\label{eq:2DTN2}.
\end{split}
\end{equation}
\end{itemize}
All indices were assigned an algebraic dimension $\chi$, except for the physical indices of the TEBD network which were assigned dimension two. Optimal contraction costs and sequences were computed in the limit of large $\chi$ (see \aref{sec:chidependency}). %
Due to the number of indices present in these expressions it is necessary to use both greek and roman alphabets twice over. The symbol $\bar{\alpha}$ (for example) consequently represents a different index to $\alpha$. No other meaning is attached to the bar, and no distinction exists between greek and roman indices, with all indices ranging from 1 to $\chi$.

All calculations proceeded to completion with the exception of the basic breadth-first algorithm and the breadth-first algorithm with restriction on outer products, neither of which terminated within a reasonable time for \Eref{eq:2DTN2}. Consequently these algorithms were only applied to the simpler tensor networks of \Erefr{eq:TEBD}{eq:2DTN1}.
Aside from these two exceptions, all algorithms successfully identified optimal-cost contraction sequences for all test networks. The optimal costs for contracting \Erefr{eq:TEBD}{eq:2DTN2} and example sequences realising these costs are given in \tref{tab:sequences}, though
it should be noted that for most networks there are multiple sequences which realise the optimal cost, and the specific sequence which is returned by an algorithm depends upon the precise order in which the search tree is explored. 
\begin{table*}[!tb]%
\caption{Optimal costs for contraction of the tensor networks given in \protect{\Erefr{eq:TEBD}{eq:2DTN2}}, and example contraction sequences which realise these costs. Sequences were computed using the pure-\MATLAB{} reference implementation of 
\prtext{\protect{\aref{sec:pseudocode}} \protect{\cite{EPAPS}}.} 
\arXtext{\protect{\aref{sec:refimp}}.}
\label{tab:sequences}}
\begin{tabular}{|c|c|c|}
\hline
Tensor & Cost & Example contraction sequence returning output as $X$\\
network & & \\
\hline\hline
3:1 1D TTN  & $4\chi^6$ & $X=((((W^{(2)} W^{(2)*})\rho)h)W^{(1)*})$\\
TEBD        & $10\chi^3+16\chi^2$ & $X=(((\lambda^{(1)}\Gamma^{(1)})(\lambda^{(2)}(\Gamma^{(2)}\lambda^{(3)})))U)$\\
3:1 1D MERA & $2\chi^8+2\chi^7+2\chi^6$ & $X=((W^{(1)} (U (h (U^* W^{(1)*})))) (W^{(2)} W^{(2)*}))$\\
9:1 2D TTN  & $4\chi^{12} + 4\chi^{10}$ & $X=((((W^{(2)} W^{(2)*}) ((W^{(3)} W^{(3)*}) ((W^{(4)} W^{(4)*}) \rho))) h) W^{(1)*})$\\
2:1 1D MERA & $2\chi^9 + 4\chi^8 + 2\chi^6 + 2\chi^5$ & $X=((W^{(1)} W^{(1)*}) (((W^{(2)} U^{(2)}) ((U^{(1)} (h U^{(1)*})) (U^{(2)*} W^{(2)*}))) (W^{(3)} W^{(3)*})))$\\
9:1 2D MERA & $3\chi^{16}+3\chi^{14}+\chi^{13}$ & 
$A=(((W^{(2)} W^{(2)*}) (V^{(3)} V^{(3)*})) (((W^{(3)} W^{(3)*}) ((W^{(4)} W^{(4)*}) (V^{(4)} V^{(4)*}))) \rho))$\\
& $~~+\chi^{12}+5\chi^{10}+5\chi^9$ & $X=(((((A (U U^*)) (V^{(2)} V^{(2)*})) (V^{(1)} V^{(1)*})) h) W^{(1)*})$ \\
4:1 2D MERA & $4\chi^{26}+2\chi^{25}+2\chi^{23}+3\chi^{22}$ & $A=((W^{(2)} W^{(2)*}) ((((((W^{(4)} W^{(4)*}) (U^{(1)} (h U^{(1)*}))) U^{(3)*}) U^{(3)}) W^{(5)*}) U^{(2)*}))$\\
& $~~+3\chi^{20}+\chi^{16}+\chi^{14}+\chi^{13}$ & $B=(((A U^{(2)}) W^{(5)}) ((W^{(6)} W^{(6)*}) ((W^{(8)} W^{(8)*}) ((W^{(9)} W^{(9)*}) (U^{(4)} U^{(4)*})))))$\\
& $~~+\chi^{12}+4\chi^8+4\chi^7$ & $X=((((B (W^{(7)} W^{(7)*})) (W^{(3)} W^{(3)*})) \rho) W^{(1)*})$\\
\hline
\end{tabular}
\end{table*}%
As the descriptions in \sref{sec:singleterm} do not specify the order in which an algorithm iterates over the members of a set $S_a$, the specific sequence which is returned may be implementation-dependent, though the optimal cost is not. %

\begin{figure}
\includegraphics[width=246.0pt]{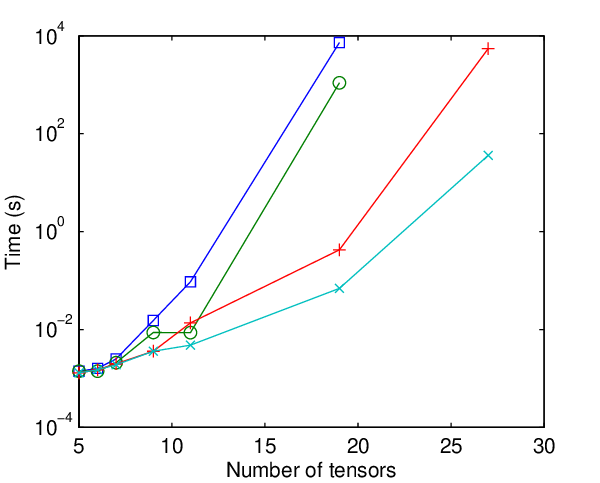}
\caption{Time to determine a guaranteed optimal contraction sequence plotted against number of tensors for different pruning algorithms. Points labelled $\square$ correspond to the exhaustive breadth-first search algorithm of \psref{sec:breadthalg} (column ``None'' in \ptref{tab:results}). Points labelled $\circ$ incorporate the restrictions on considered outer products described in \psref{sec:restrictops} (column ``OP'' in \ptref{tab:results}). Points labelled $+$ correspond to the cheapest-first variant on the breadth-first algorithm, described in \psref{sec:costcap} (column ``$\mu_\mrm{cap}$'' in \ptref{tab:results}). Points labelled $\times$ correspond to the cheapest-first variant on the breadth-first algorithm with restrictions on outer products (column ``OP~\&~$\mu_\mrm{cap}$'' in \ptref{tab:results}).\label{fig:results}}
\end{figure}

The times taken by each algorithm to identify each optimal cost along with a corresponding contraction sequence are given in \tref{tab:results} and plotted in \fref{fig:results}. These results were computed 
using a Dell workstation with a 3.6GHz Intel Xeon processor and 48Gb of 1333MHz DDR3 RAM, %
though near-identical performance was obtained on an early-2011 Macbook Pro with a 2.2GHz Intel i7 processor and 8Gb of 1333MHz DDR3 RAM.%
With cost capping enabled, memory poses no significant constraint even for networks as large as the 4:1~2D~MERA (27 tensors), and from the lack of dependence on processor speed we infer that performance of the algorithm is memory bandwidth limited.
\begin{table}[bp]%
\caption{Time in seconds to find a guaranteed optimal contraction sequence for the tensor networks of Figs.~\protect{\ref{fig:31MERA}}, \protect{\ref{fig:1DTNs}}, and \protect{\ref{fig:2DTNs}}
using a breadth-first search. The first column of results corresponds to the basic algorithm of \psref{sec:breadthalg}, ``$\xicap$'' indicates application of the cheapest-first algorithm described in \psref{sec:costcap}, and ``OP'' indicates restriction of allowed contractions creating or involving outer products according to \psref{sec:restrictops}.\label{tab:results}}
~\\\begin{tabular}{|c|c|cccc|}
\hline
Tensor&Number of&\multicolumn{4}{c|}{Pruning techniques:}\\ %
network&tensors& None & OP & $\xicap$ & OP \& $\xicap$\\ \hline \hline
3:1 1D TTN  & 5& 0.0014  & 0.0014 & 0.0013 & 0.0013 \\
TEBD        & 6& 0.0016  & 0.0014 & 0.0015 & 0.0015 \\
3:1 1D MERA & 7& 0.0025  & 0.0021 & 0.0020 & 0.0019 \\
9:1 2D TTN  & 9& 0.0152  & 0.0087 & 0.0036 & 0.0036 \\
2:1 1D MERA &11& 0.0946  & 0.0086 & 0.0136 & 0.0048 \\
9:1 2D MERA &19&    7298 & 1096   & 0.423  & 0.069 \\
4:1 2D MERA &27& \footnote{Aborted: Insufficient memory to perform calculation without swapping to disk.\label{tablenote1}} & $^\mathrm{\protect{\ref{tablenote1}}}$ &   5507 & 36 \\ \hline
\end{tabular}
\end{table}%

The software environment on the workstation comprised \MATLAB{} release 2012b paired with Gnu~C++~4.7.2-2ubuntu1
for compilation of the C++ component, while that on the laptop consisted of \MATLAB{} release 2011a paired with with Apple~Xcode~5.0.2. 

From these results the enhanced pruning algorithms of Secs.~\ref{sec:costcap} and \ref{sec:restrictops} are each seen to be  capable of yielding performance increases of up to several orders of magnitude, dependent upon the size and structure of the tensor network being analysed. 

It is recognised that this performance benefit is dependent upon the connectivity of the network being studied, and that (for example) all benefits of \sref{sec:restrictops} will disappear in the limit of a fully-connected network (where every tensor shares an index with every other tensor). However, tensor networks in condensed matter physics are typically much more sparsely connected than the fully-connected network, and the MERA networks used as examples here represent a comparatively highly-connected example. Furthermore, the benefit persisted even for networks with as few as five or six tensors. %
The pruning algorithms of \sref{sec:beyond} %
are therefore anticipated to be of substantial practical benefit. %

\section{Discussion}

\subsection{Faster, better Ansatz development and implementation}

This paper has presented a detailed description (and a reference implementation) of an improved algorithm for determining an optimal sequence for the contraction of a tensor network (or, in the language of quantum chemistry, a single term in a tensor expression), along with the associated computational cost. The benefit of such automated algorithms is well-established in quantum chemistry, with the ability to reduce the development time for novel Ans\"atze by several years \cite{auer2006}. Their adoption in condensed matter physics has been slower primarily due to the larger tensor networks involved, and the NP-hard scaling of such algorithms with the number of tensors. The algorithm presented in this paper is demonstrated to be faster than previously-favoured exhaustive search algorithms for networks of five or more tensors, and is faster by several orders of magnitude for some of the more complex tensor networks encountered in condensed matter physics, making the automated determination of optimal contraction sequences both viable and attractive.

While it could be argued that the tensor network Ans\"atze of condensed matter physics are only just beginning to show the level of complexity where automated determination of optimal contraction sequences is necessary, the advantages of doing so have already been demonstrated: In \rcite{pfeifer2013}, one of the authors applied an earlier implementation of some of the techniques described in \sref{sec:beyond} to the 4:1 2D MERA. At the time, assuming all indices to have a dimension of $\chi$, the best-known contraction sequences for this Ansatz scaled as $\OO{\chi^{28}}$ \cite{evenbly2009b}; the automated search revealed a contraction sequence for a cost of $\OO{\chi^{26}}$.

Automated determination of optimal contraction sequences relieves researchers from this lengthy and tiresome task, and may on occasion lead to implementations of algorithms which are more efficient than those discovered by hand. As increasingly sophisticated Ans\"atze are proposed, the need for automated search algorithms to determine optimal tensor contraction sequences in condensed matter physics will only increase.

\subsection{New possibilities in condensed matter and quantum gravity}

To date, tensor network algorithms in condensed matter physics have employed fixed networks of tensors, where the coefficients of the tensors may be updated but the structure of the network itself remains unchanged, arguably due to the requirement that optimal contraction sequences for these tensor networks be hard-coded into the implementing software at the time of programming. With the advent of an efficient algorithm to find the optimal contraction sequence for an arbitrary network of tensors at runtime, algorithms which involve a dynamically evolving tensor network become feasible. One could, for example, propose an initial Ansatz consisting of a very highly connected tensor network with comparatively low index dimensions, and then perform variational optimisation while allowing the dimensions of these indices to vary according to the spectra of appropriately-chosen Schmidt decompositions. 
By increasing bond dimensions where the spectra are flat and decreasing them where the spectra decay more rapidly, a tensor network may evolve to represent the entanglement structure of the state under study. 
(A simple example of this is automated reduction of index dimension in DMRG for weakly-correlated states, though in this instance the optimal contraction sequence generally remains unchanged.) 
Use of a software algorithm to determine optimal contraction sequences (and to prohibit changes to the network whose cost is too great) guarantees that the resulting tensor network Ans\"atze are contracted as efficiently as possible. The concept becomes especially attractive when one considers results presented in \rcite{evenbly2014} showing how an $\OO{N}$ speed-up may be obtained for variational optimisation of entirely non-symmetric tensor networks, making self-adaptive networks a potentially viable proposition even in the absence of any obvious physical (e.g.~spatial) symmetry.

The Netcon algorithm may also be of interest in the numerical study of loop quantum gravity. In loop quantum gravity, space is represented as a \emph{spin network}, which may be understood as a dynamically evolving network of SU(2)-symmetric tensors,
and calculation of observables with respect to a given spin network necessarily involves the contraction of a large tensor network. Further, in evolving from a state A to a state B, multiple tensor contractions may take place, and %
topological equivalence may give some freedom as to the order in which these contractions are performed. In both situations there is a need to determine the optimal sequence with which to perform tensor contractions, and therefore there is a role for an algorithm such as Netcon. Further, the proven ability of Netcon to analyze networks involving a couple of dozen tensors with only modest requirements in both memory and computation time make the algorithm extremely well-suited to calculations on the sort of scales at which current numerical simulations are likely to take place.

\acknowledgments

R.N.C.P. thanks the Ontario Ministry of Research and Innovation Early Researcher Awards for financial support. Research at Perimeter Institute is supported by the Government of Canada through Industry Canada and by the Province of Ontario through the Ministry of Research and Innovation. %
F.V. acknowledges support from the European Research Council (Belgium), %
the Austrian Science Fund (ViCoM), the SFB FoQuS, and the Swiss Institute for Quality Standards.

Note on authorship: The cost-capping techniques and restrictions on outer products described in \sref{sec:beyond} were originally developed by R.N.C.P. in the context of a depth-first search as described in \sref{sec:depthalg}. J.H. and F.V. then implemented cost capping in the context of a breadth-first search as per \sref{sec:breadthalg}, and demonstrated that this yielded a further substantial improvement in performance. The authors combined their developments and proceeded collaboratively to publication.

\appendix

\section{Matrix chain multiplication\label{sec:matmult}}

The matrix chain multiplication problem is a well known problem in linear algebra, in which one must determine the least number of mathematical operations required to evaluate the product of a series of varyingly-sized matrices.
For example, given the calculation
\begin{equation}
D_{ij} = A_{ik}\times B_{kl}\times C_{lj},
\end{equation}
where indices $i$, $j$, $k$, and $l$ range from 1 to $\xi_i$, $\xi_j$, $\xi_k$, and $\xi_l$ respectively (and $\xi_x$ is termed the \emph{dimension} of index $x$),
one may first multiply $A$ by $B$ then multiply $(AB)$ by $C$ at a cost of $\xi_i\xi_k\xi_l+\xi_i\xi_l\xi_j$ operations, or first multiply $B$ by $C$ then multiply $A$ by $(BC)$ at a cost of $\xi_k\xi_l\xi_j+\xi_i\xi_k\xi_j$ operations, where each operation comprises one floating point multiplication followed by one floating point addition.

A matrix may only ever be multiplied by one of at most two immediate neighbours,\footnote{Indeed, the constraints of \psref{sec:restrictops} confirm that outer products need never be considered when looking for the optimal contraction of a matrix chain.} and as a consequence an optimal sequence of pairwise matrix products may always be found in polynomial time scaling as $\OO{n^3}$ using dynamic programming techniques \cite{cormen2001}. 
For networks where a tensor may share indices with more than two immediate neighbours, however, this reduction to polynomial time breaks down, and as seen in \sref{sec:restrictops} the optimal contraction sequence may then even require performing the outer product between two or more tensors which do not share a common index. 
(Constraints derived in \sref{sec:tensorsinOP} show that there must exist at least one tensor sharing indices with three neighbours for this to be necessary, and so outer products are never required for the matrix chain.) 
One is therefore forced to consider all pairwise contractions regardless of whether or not any indices are shared by the participating tensors, and the problem of determining an optimal contraction sequence for a general tensor network has been shown to be NP-hard \cite{lam1997}. 

For matrix chain multiplication, an even faster algorithm is known which returns an optimal sequence in time scaling as $\OO{n\log{n}}$ \cite{hu1982,hu1984}. However, no extension of this algorithm to more general tensor networks has yet been proposed.

\section{Preferential nature of pairwise contractions\label{sec:pairwise}}

In \sref{sec:TNAs} it was stated that an optimal contraction sequence for a tensor network may always be realised as a series of pairwise contractions. To see this, consider the contraction of three tensors, $A$, $B$, and $C$, to yield a single tensor~$D$. Let $\xi_{AB}$ denote the product of the dimensions of all indices on tensor~$A$ which connect to tensor~$B$, and similarly for $\xi_{AC}$ and $\xi_{BC}$. Let $\xi_A$ denote the product of all indices on tensor~$A$ which do not connect to either $B$ or $C$, and similarly for $\xi_B$ and $\xi_C$. The dimension of a set containing no indices, is always one. Contracting these three tensors as a single process involves a cost of
\begin{equation}
2\xi_A\xi_B\xi_C\xi_{AB}\xi_{AC}\xi_{BC}.\label{eq:triplecontract}
\end{equation}
For example, for the contraction
\begin{equation}
D^{a}_{e}=A^{abc}B_{bd}C^{d}_{ce},
\end{equation}
we have
\begin{equation}
\begin{split}
\xi_A&=|a|,\\ \xi_B&=1,\\ \xi_C&=|e|,\\ \xi_{AB}&=|b|,\\
\xi_{AC}&=|c|, \\\mrm{and}~\xi_{BC}&=|d|.
\end{split}
\end{equation}
For each element of $D^\alpha_\epsilon$ it is necessary to sum over $\xi_{AB}\xi_{AC}\xi_{BC}$ different contributions (corresponding to the enumeration of indices $b$, $c$, and $d$), each involving two multiplications, and there are then $\xi_A\xi_B\xi_C$ entries in $D^a_e$, for the total number of multiplication operations given in \Eref{eq:triplecontract}.
In contrast, pairwise contraction may be achieved by any of the sequences $((AB)C)$, $((AC)B)$, or $((BC)A)$, where $((XY)Z)$ means ``contract tensor~$X$ with tensor~$Y$, then contract the result with tensor~$Z$''. The sequence $((AB)C)$ is readily seen to attract a total cost of
\begin{equation}
\xi_A\xi_B\xi_{AB}\xi_{AC}\xi_{BC} + \xi_A\xi_B\xi_C\xi_{AC}\xi_{BC}\label{eq:doublecontract}
\end{equation}
multiplication operations, with those for $((AC)B)$ and $((BC)A)$ being achieved by the relevant label permutations. Since all parameters in \Erefs{eq:triplecontract}{eq:doublecontract} take a value greater than or equal to one, the value of \Eref{eq:doublecontract} is always less than or equal to that of \Eref{eq:triplecontract}. The argument extends directly to contraction of four or more tensors, with the cost of sequential pairwise contraction continuing to always be less than or equal to that of more complicated contractions, and consequently for any tensor network it is always possible to identify a minimum-cost contraction sequence in which all contractions proceed in a pairwise fashion. Note that no assumption has been made about the values of the index dimensions, and thus this result holds even for contraction sequences involving outer products, which may be equated with contraction over indices of dimension one.

\section{Exclusion of sequences $((AB)(DE))$ and $(((AB)E)D)$ for $\xd=\xc$ in \pfref{fig:(AB)DE}(c)\label{sec:tensorswithOP_noAB}}

In \sref{sec:tensorswithOP} we showed that to find an optimal contraction sequence we only need consider outer products where the resulting object is contracted with a fundamental tensor (supplied as input to the algorithm), or with a composite tensor (denoted $C$) where the final step in the lowest-cost construction of this tensor necessarily takes the form of contracting $D$ with $E$ in \fref{fig:(AB)DE}(c). It was also shown that indices $\xa$, $\xc$, and $\xd$ must satisfy the constraints
\begin{align}
\xa&<\xc\tag{\ref{eq:Cconstr2}}\\
\xd&\leq\xc,\tag{\ref{eq:Cconstr3}}
\end{align}
and that where multiple lowest-cost constructions of tensor~$C$ exist, all must be consistent with \fref{fig:(AB)DE}(c) and it is the smallest value of $\xc$ obtained which must satisfy \Erefs{eq:Cconstr2}{eq:Cconstr3}. We now show that \Eref{eq:Cconstr3} may be tightened to
\begin{equation}
\xd<\xc\tag{\ref{eq:Cconstr1}}.
\end{equation}

\begin{figure}
\includegraphics[width=246.0pt]{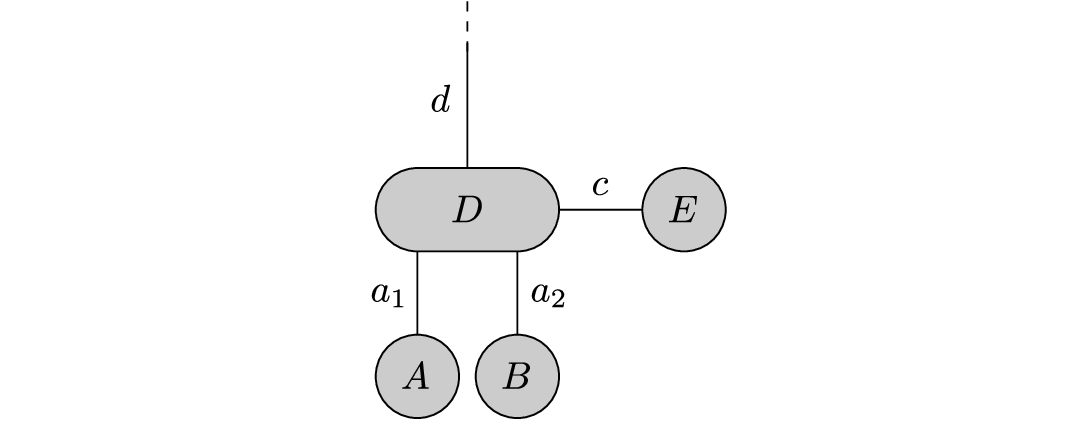}
\caption{Network $\mc{N}_{ABDE}$ of \pfref{fig:(AB)DE}(b) prior to contraction of tensor~$A$ with tensor~$B$.\label{fig:ABDE}}
\end{figure}

In \sref{sec:tensorswithOP} we considered the contraction of a tensor~$(AB)$, formed by performing an outer product between tensors~$A$ and~$B$, with tensors~$D$ and~$E$ which make up composite tensor~$C$, under the assumption that contraction $(AB)$ was always performed first. We then asked when it was preferable to contract tensor~$D$ with tensor~$E$, thus obtaining tensor~$C$, rather than contracting tensor~$(AB)$ directly with either~$D$ or~$E$. We now relax this setup to consider the network of four tensors, $A$, $B$, $D$, and $E$, prior to any contraction but subject to the constraint that $\xc=\xd$. This network is shown in \fref{fig:ABDE}, and \Eref{eq:Cconstr2} now corresponds to
\begin{equation}
\xi_{a_1}\xi_{a_2}<\xc\label{eq:ABDECconstr1}
\end{equation}
where $\xi_{a_1}\xi_{a_2}=\xa$.

From \sref{sec:tensorswithOP} we know that when $\xc=\xd$ the costs of sequences $((AB)(DE))$ and $(((AB)E)D)$ are equal and are given by
\begin{equation}
\xi_{a_1}\xi_{a_2} + \xi_{a_1}\xi_{a_2}\xc^2 + \xi_{a_1}\xi_{a_2}\xc.
\end{equation}
We now compare this with the cost of sequence $(((AE)D)B)$, which evaluates to
\begin{equation}
\xi_{a_1}\xc+\xi_{a_1}\xi_{a_2}\xc^2+\xi_{a_2}\xc. 
\end{equation}
For $((AB)(DE))$ or $(((AB)E)D)$ to be cheaper than $(((AE)D)B)$ requires
\begin{equation}
\begin{split}
\xi_{a_1}\xi_{a_2} + \xi_{a_1}\xi_{a_2}\xc^2 + \xi_{a_1}\xi_{a_2}\xc &< \xi_{a_1}\xc+\xi_{a_1}\xi_{a_2}\xc^2+\xi_{a_2}\xc\\
\xi_{a_1}\xi_{a_2} &< (\xi_{a_1}+\xi_{a_2}-\xi_{a_1}\xi_{a_2})\xc\\
\Rightarrow 0 &< \xi_{a_1}+\xi_{a_2}-\xi_{a_1}\xi_{a_2}.
\end{split}
\end{equation}
Non-disjointness of the network requires $\xi_{a_1}\geq2$ and $\xi_{a_2}\geq2$ so this condition is never satisfied, and we may be sure of identifying an optimal contraction sequence without considering either $((AB)(DE))$ or $(((AB)E)D)$. Consequently it is entirely acceptable that $(((AB)E)D)$ be excluded by \Eref{eq:OPconds} and $((AB)(DE))$ by \Eref{eq:Cconstr1}. Meanwhile, for sequence $(((AE)D)B)$ %
the constraints of \Eref{eq:OPconds} %
become
\begin{equation}
\xi_{a_1}\xc>\xi_{a_2}\qquad\xi_{a_1}\xc>\xc
\end{equation}
and are automatically satisfied by virtue of \Eref{eq:ABDECconstr1} and $\xi_{a_1}\geq2$.

\section{Pseudocode implementation of \psref{sec:beyond}\label{sec:pseudocode}}

This Appendix presents a pseudocode implementation of the pruning algorithms described in \sref{sec:costcap} and \sref{sec:restrictops}. The constraints on tensor $C$ of \sref{sec:tensorswithOP} are realised by constructing a list (called $L$) containing the sets of index labels appearing on admissible tensors, along with the index dimension corresponding to $\xc=|E|$ in \fref{fig:(AB)DE}(c). 
Whenever the best-known contraction sequence for a tensor $C$ ends with a contraction $(DE)$ consistent with the structure shown in \fref{fig:(AB)DE}(c), its details are added to the list. However, if a sequence is subsequently identified of equal or lower cost which does not satisfy this construction, the details of $C$ are removed from the list again. If this is going to happen, it always does so before the next increment of $\mu_\mrm{cap}$. Additions to the list are therefore made provisionally, to be confirmed once a pass with a given value of $\mu_\mrm{cap}$ is complete,
and 
outer products are only performed if there exists at least one non-provisional entry in the list which is consistent with the requirements for tensor~$C$ given in \sref{sec:tensorswithOP}.

If any provisional entries are present on the list when $\mu_\mrm{cap}$ is due to be increased, then these entries are confirmed and another pass is performed at the same value of $\mu_\mrm{cap}$ to resolve the new contractions which are made possible by the additional non-provisional entries in the list. Waiting until the initial pass at cost $\mu_\mrm{cap}$ is complete before performing the newly-available outer products means that if a provisional entry is added to the list but is then subsequently deleted, or its constraints are tightened, exploration of the corresponding unnecessary branches of the search tree can largely be avoided.\footnote{This avoidance may not be perfect; consider the following situation: Let a tensor~\protect{$X$}, consistent with \protect{$(DE)$} in \pfref{fig:(AB)DE}(c), be added to list~\protect{$L$} and subsequently become non-provisional. Let another tensor~\protect{$Y$} be added to list~\protect{$L$} and become non-provisional at the same time or later than~\protect{$X$}. If addition of tensor~\protect{$Y$} to list~\protect{$L$} makes it possible to construct~\protect{$X$} for equal-best or better cost than previously known, and in a manner either inconsistent with \pfref{fig:(AB)DE}(c) or yielding a lower value of \protect{$\xc$} than previously obtained, then it is possible that time could have been wasted on the performance of unnecessary contractions \protect{$(AB)$} and \protect{$((AB)X)$}, where \protect{$(AB)$} is an outer product, prior to finding the new sequence for~\protect{$X$} and implementing the improved constraints which that sequence implies. In practice, this scenario has yet to be observed.}

As per \sref{sec:singleterm} it is assumed that all indices are of dimension~2 or higher, and that the tensor network is non-disjoint.\arXtext{ For discussion of how to incorporate disjoint networks or networks including indices of dimension~1, see \aref{sec:subnets}.}\prtext{ A discussion of how this algorithm may be extended to disjoint networks and networks including indices of dimension~1 may be found in the instructions accompanying the reference implementation \cite{EPAPS}.}

\begin{enumerate}
\item[] $\!\!\!\!\!\!\!$\textsc{Algorithm: Netcon}
\item \rule{0pt}{3.7ex}Let $S_1=\{T_1,\ldots,T_n\}$ be the set of $n$ tensors which make up network $\mc{N}$.
\item Flag each tensor in $S_1$ as ``old''.
\item Let $\{S_i~|~i\in\mbb{Z},~2\leq i\leq n\}$, be empty sets.
\item Let $\mu_\mrm{cap}=1$, let $\mu_\mrm{old}=0$, let $\mu_\mrm{next}=\infty$, and let $\xi_\mrm{min}$ be the dimension of the smallest index on any tensor.
\item Let $L$ be an empty list whose entries $l\in L$ each take the form $l\equiv\{I_l,\xi_l,f_l\}$ where $I_l$ is a list of indices, $\xi_l$ is a tensor dimension, and $f_l$ is a numerical flag. For each $T_i$ in turn, perform sub-algorithm \textsc{AddToList[$T_i$,$-$]}.
\item Assign flag 0 (``old entry'') to all entries in $L$.
\item While $S_n$ is empty:
\begin{enumerate}
\item Let $c$ be a counter running from 2 to $n$.\\For each value of $c$,\\and each pair of sets $S_d$, $S_{c-d}$, $1\leq d\leq \lfloor \frac{c}{2}\rfloor$,\\and each $T_a\in S_d$, $T_b\in S_{c-d}$ such that each element of $S_1$ appears at most once in $\fuse{T_a}{T_b}$:\label{alg:iterate}
\begin{enumerate}
\item If $T_a$ and $T_b$ share no common indices:
\begin{enumerate}
\item If $\mu_\mrm{old}\not=\mu_\mrm{cap}$, and there exists an entry $l$ in $L$ with flag $f_l=0$ (``old'') for which $I_l$ contains all indices on $T_a$ and $T_b$ and $\xi_l\geq |T_a||T_b|$, advance to step~\ref{alg:costit}.
\item If $\mu_\mrm{old}=\mu_\mrm{cap}$, and there exists an entry $l$ in $L$ with flag $f_l=1$ (``new'') for which $I_l$ contains all indices on $T_a$ and $T_b$ and $\xi_l\geq |T_a||T_b|$, advance to step~\ref{alg:costit}.
\item If $\mu_\mrm{old}=\mu_\mrm{cap}$, either of $T_a$ or $T_b$ is flagged as ``new'', and there exists an entry $l$ in $L$ with flag $f_l=0$ for which $I_l$ contains all indices on $T_a$ and $T_b$ and $\xi_l\geq |T_a||T_b|$, advance to step~\ref{alg:costit}.
\item Otherwise, return to step~\ref{alg:iterate} and select the next pair $\{T_a,T_b\}$.
\end{enumerate}
\item If either $T_a$ or $T_b$ is the result of an outer product, and contraction of $T_a$ with $T_b$ is not an outer product:
\begin{enumerate}
\item If both $T_a$ and $T_b$ are outer products, this violates the constraints of \sref{sec:tensorswithOP}. Return to step~\ref{alg:iterate} and select the next pair $\{T_a,T_b\}$.
\item Let $Y$ be the member of $\{T_a,T_b\}$ which is an outer product, and $X$ be the member of $\{T_a,T_b\}$ which is not an outer product. If $X$ is not in $S_1$ and does not satisfy the form mandated in \fref{fig:(AB)DE}(c), return to step~\ref{alg:iterate} and select the next pair $\{T_a,T_b\}$.
\item If $X$ and $Y$ do not satisfy \Eref{eq:OPconds}, or $X$ is not in $S_1$ and $X$ and $Y$ do not satisfy \Erefs{eq:Cconstr2}{eq:Cconstr1}, return to step~\ref{alg:iterate} and select the next pair $\{T_a,T_b\}$.
\end{enumerate}
\item If either $T_a$ or $T_b$ is the result of an outer product, and contraction of $T_a$ with $T_b$ is an outer product, check that it satisfies \Eref{eq:OPconds2*}. Otherwise, return to step~\ref{alg:iterate} and select the next pair $\{T_a,T_b\}$.
\item Let $\mu=\ttt{cost}[\fuse{T_a}{T_b}]$.\label{alg:costit}
\item Where $T_a$ and/or $T_b$ do not belong to $S_1$, add to $\mu$ the previously-determined cost of constructing $T_a$ and/or $T_b$ as appropriate.
\item If either $T_a$ or $T_b$ is flagged as ``new'', let $\mu_0=0$. Otherwise, let $\mu_0=\mu_\mrm{old}$.
\item If $\mu>\mu_\mrm{cap}$ and $\mu<\mu_\mrm{next}$, let $\mu_\mrm{next}=\mu$.
\item If $\mu_0<\mu\leq\mu_\mrm{cap}$: 
\begin{enumerate}
\item Let the contraction sequence $\mc{Q}$ for constructing this object be written $\mc{Q}=\fuse{T_a}{T_b}$. Where $T_a$ and/or $T_b$ do not belong to $S_1$, the best-known contraction sequences for $T_a$ and $T_b$ will have been previously recorded: In $\mc{Q}$, replace each appearance of $T_a$ and/or $T_b$ with the corresponding best-known contraction sequences.
\item If no object corresponding to $\fuse{T_a}{T_b}$ has yet been created in $S_c$, create it. Otherwise, locate the object in $S_c$ which corresponds to $\fuse{T_a}{T_b}$.
\item If this is the first known sequence for constructing this object, or $\mu$ is cheaper than any previously-known cost for constructing this object:
\begin{itemize}
\item[I.] Record the cost $\mu$ and the associated contraction sequence $\mc{Q}$ against this object.
\item[II.] Flag the object as ``new''. 
\item[III.] If $T_a$ and $T_b$ have structures permitting them to be identified with $D$ and $E$ in \fref{fig:(AB)DE}(c) and their dimensions satisfy \Eref{eq:cubevsD}, perform sub-algorithm \textsc{AddToList[$\fuse{T_a}{T_b}$,$\mc{Q}$]}. Otherwise, if this is not the first known sequence for constructing this object, perform sub-algorithm \textsc{RemoveFromList[$\fuse{T_a}{T_b}$]}.
\end{itemize}
\item If this is not the first known sequence for constructing this object, $\mu$ is equal to the best previously-known cost for constructing this object, and the final contraction in the previous best-known sequence for this object can be identified with $(DE)$ in \fref{fig:(AB)DE}(c):
\begin{itemize}
\item[I.] If $T_a$ and $T_b$ have structures permitting them to be identified with $D$ and $E$ in \fref{fig:(AB)DE}(c), and the value of $\xc$ in \fref{fig:(AB)DE}(c) for sequence~$\mc{Q}$ is lower than that of the previous best recorded sequence for $(T_aT_b)$ but still satisfies \Eref{eq:cubevsD}, replace that sequence with $\mc{Q}$ then perform sub-algorithm \textsc{UpdateList[$\fuse{T_a}{T_b}$,$\mc{Q}$]}.
\item[II.] If $T_a$ and $T_b$ have structures permitting them to be identified with $D$ and $E$ in \fref{fig:(AB)DE}(c), and the value of $\xc$ in \fref{fig:(AB)DE}(c) for sequence~$\mc{Q}$ is lower than that of the previous best recorded sequence for $(T_aT_b)$ and does not satisfy \Eref{eq:cubevsD}, replace that sequence with $\mc{Q}$ then perform sub-algorithm \textsc{RemoveFromList[$\fuse{T_a}{T_b}$,$\mc{Q}$]}.
\item[III.] If $T_a$ and $T_b$ do not have structures permitting them to be identified with $D$ and $E$ in \fref{fig:(AB)DE}(c), replace the previous best recorded sequence for $(T_aT_b)$ with $\mc{Q}$ and perform sub-algorithm \textsc{RemoveFromList[$\fuse{T_a}{T_b}$]}.
\end{itemize}
\item If $\fuse{T_a}{T_b}\in S_n$, set $\mu_\mrm{cap}=\mu$. (Solution found; no need to consider any sequences more expensive than this.)
\end{enumerate}
\end{enumerate}
\item Let $\mu_\mrm{old}=\mu_\mrm{cap}$.
\item If no entries in $L$ are flagged 2 (``provisional''), set $\mu_\mrm{cap}$ equal to the larger of $\mu_\mrm{next}$ and $\xi_\mrm{min}\mu_\mrm{cap}$, then set $\mu_\mrm{next}=\infty$.
\item Flag all tensors in all $S_i$ as ``old''.
\item Confirm provisional additions to $L$: Decrease the flags on all entries in $L$ by 1, to a minimum of 0.
\item Remove redundant entries in $L$: For each entry $l\in L$ and for each entry $l'\in \{L~|~f_{l'}=1,~l'\not=l\}$, if $I_l\subset I_{l'}$ and $\xi_l\leq \xi_{l'}$, delete entry $l$ from $L$.
\end{enumerate}
\item The optimal cost $\mu_\mrm{best}$ and a sequence $\mc{Q}_\mrm{best}$ which realises this are recorded against the only element in $S_n$.
\end{enumerate}
\begin{enumerate}
\item[] $\!\!\!\!\!\!\!$\textsc{SubAlgorithm: AddToList[$X$,$\mc{Q}$]}
\item \rule{0pt}{3.7ex}If $X$ is in $S_1$, let $\xi=\infty$. Otherwise, considering the final step in sequence~$\mc{Q}$, let $\xi$ be the dimension of the tensor which corresponds to $E$ in \fref{fig:(AB)DE}(c).
\item An entry in $L$ takes the form $l=\{I_l,\xi_l,f_l\}$ where $I_l$ is a list of indices, $\xi_l$ is a tensor dimension, and $f_l$ is a flag. Stepping through all non-provisional entries $\{l\in L~|~f_l\not=2\}$:
\begin{enumerate}
\item If indices on $X$ exactly match $I_l$:
\begin{enumerate}
\item Set $\xi_l=\xi$.
\item If this corresponds to an increase in the value of $\xi_l$, constraints have been relaxed. Set $f_l=2$ to ensure another pass at the same value of $\mu_\mrm{cap}$.
\item Terminate subalgorithm \textsc{AddToList}.
\end{enumerate}
\item If $f_l\not=2$, all indices on $X$ appear in $I_l$, and $\xi\leq \xi_l$:
\begin{enumerate}
\item An entry for $X$ would be redundant. One may previously have been created in $L$ due to an earlier sequence with a higher value of $\xi$. If so, this is now superceded by the present value of $\xi$ and so is also redundant. Perform subalgorithm \textsc{RemoveFromList[$X$]}.
\item Terminate subalgorithm \textsc{AddToList}.
\end{enumerate}
\end{enumerate}
\item If an entry for $X$ may previously have been created in $L$ with higher $\xi$, step through all provisional entries $\{l\in L~|~f_l=2\}$. If an entry $l$ is found for which $I_l$ exactly matches the indices on $X$:
\begin{enumerate}
\item Set $\xi_l=\xi$.
\item Terminate subalgorithm \textsc{AddToList}.
\end{enumerate}
\item Add a new entry $l'$ to $L$, where $I_{l'}$ is a list of all indices appearing on $X$, $\xi_{l'}=\xi$, and $f_{l'}=2$ (``provisional'').
\end{enumerate}
\begin{enumerate}
\item[] $\!\!\!\!\!\!\!$\textsc{SubAlgorithm: UpdateList[$X$,$\mc{Q}$]}
\item \rule{0pt}{3.7ex}Considering the final step in sequence~$\mc{Q}$, let $\xi$ be the dimension of the tensor which corresponds to $E$ in \fref{fig:(AB)DE}(c).
\item An entry in $L$ takes the form $l=\{I_l,\xi_l,f_l\}$ where $I_l$ is a list of indices, $\xi_l$ is a tensor dimension, and $f_l$ is a flag. Stepping through all entries $\{l\in L\}$, if $I_l$ exactly matches the indices on~$X$:
\begin{enumerate}
\item Set $\xi_l=\xi$.
\item Terminate subalgorithm \textsc{UpdateList}.
\end{enumerate}
\end{enumerate}
\begin{enumerate}
\item[] $\!\!\!\!\!\!\!$\textsc{SubAlgorithm: RemoveFromList[$X$]}
\item \rule{0pt}{3.7ex}An entry in $L$ takes the form $l=\{I_l,\xi_l,f_l\}$ where $I_l$ is a list of indices, $\xi_l$ is a tensor dimension, and $f_l$ is a flag. Stepping through all entries $l\in L$, if the indices on $X$ exactly match $I_l$:
\begin{enumerate}
\item Delete entry $l$ from $L$.
\item Terminate subalgorithm \textsc{RemoveFromList}.
\end{enumerate}
\end{enumerate}

\begin{figure}
\raisebox{-52pt}{\includegraphics[width=123.0pt]{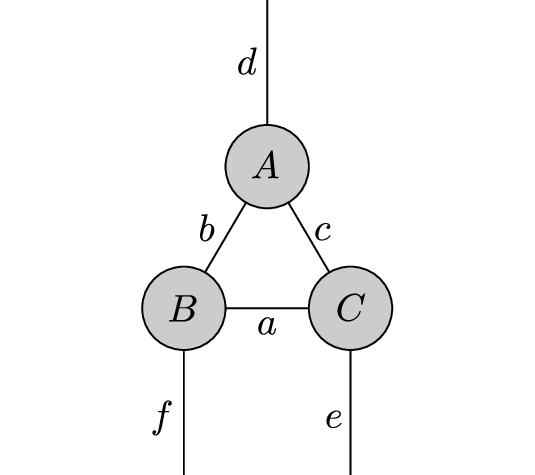}}
\begin{tabular}{|c|c|}
\hline
Index label & Dimension\\
\hline\hline
a & $6\chi$\\
b & $\chi$\\
c & $\chi^2$\\
d & $6\chi$\\
e & $\chi$\\
f & $\chi^2$\\\hline
\end{tabular}~~~~~~~~~
\caption{A simple network for which the optimal contraction sequence varies with the value of $\chi$.\label{fig:threetensors}}
\end{figure}%
\arXtext{\section{Reference implementation\label{sec:refimp}}}

\arXtext{\subsection{Obtaining and compiling}}

A reference implementation of the Netcon algorithm\arXtext{ of \aref{sec:pseudocode}}, written in \MATLAB{} and C++, may be\prtext{ found in the %
supplemental material associated with this paper \cite{EPAPS}.}\arXtext{ obtained as follows: While viewing the abstract page for this paper on arXiv.org, select ``Download $>$ Other formats'', then ``Download source''. Save the download with extension \texttt{.tar}. On expanding the resulting archive, the files comprising an implementation of the %
algorithm are \texttt{netcon.m} and \texttt{netcon\_nondisj\_cpp.cpp}.}
The implementation requires \MATLAB{}~2011a or above, and performance may be enhanced using of a compatible C++ compiler. It has been tested under \MATLAB{}~2011a with %
Apple~XCode~5.0.2 and \MATLAB{}~2012b with Gnu~C++~4.7.2-2ubuntu1. \prtext{The network contraction sequences output by the reference implementation are fully compatible with the tensor network contraction packages \ttt{ncon()} and \ttt{multienv()} of Refs.~\onlinecite{pfeifer2014} and~\onlinecite{evenbly2014} respectively.}\arXtext{(Note: The arXiv download page states that the source will be packaged in \texttt{.tar.gz} format. This is incorrect; the download is packaged in \texttt{.tar} format only.)}

\arXtext{
The reference implementation %
comprises a \MATLAB{} function \ttt{netcon()}, contained in the file \ttt{netcon.m}, which may be invoked from the \MATLAB{} command line. The performance of this reference implementation may be optionally improved by compiling the C++ component provided in \ttt{netcon\_nondisj\_cpp.cpp}. Assuming that an appropriate compiler has been installed and configured using the \MATLAB{} command \ttt{mex~-setup}, the C++ component may be compiled by typing
\begin{enumerate}
\item[] \ttt{mex netcon\_nondisj\_cpp.cpp}
\end{enumerate}
at the \MATLAB{} command prompt.

Contraction sequences output by \ttt{netcon()} are fully compatible with the tensor network contraction packages \ttt{ncon()} and \ttt{multienv()} of Refs.~\onlinecite{pfeifer2014} and~\onlinecite{evenbly2014} respectively.

\varsubsection{Invocation}

Invocation of the algorithm is via the \MATLAB{} command
\begin{align*}
\ttt{[sequence cost] = netcon(legLinks,verbosity,}\\\ttt{costType,muCap,allowOPs,legCosts);}
\end{align*}
and takes between one and six input parameters, as follows:

\ttt{legLinks}: This parameter describes the tensor network for which an optimal contraction sequence is sought. To construct \ttt{legLinks}, first draw the tensor network using the customary graphical notation (summarised in \S{}1.2 of Ref.~\onlinecite{pfeifer2011a}), with each tensor being represented by a shape, each summed index by a line connecting the two tensors on which it appears, and each unsummed index by a line with one free end and the other end attached to the tensor on which it appears\arXtext{ [for example,~\fref{fig:31MERA} is a graphical representation of Eq.~\eref{eq:31MERA}]}.
Next, label each summed index with a unique positive integer, and each unsummed index with a unique negative integer, descending consecutively from $-1$ (e.g.~\fref{fig:labelled31MERA}). 
\begin{figure}
\includegraphics[width=246.0pt]{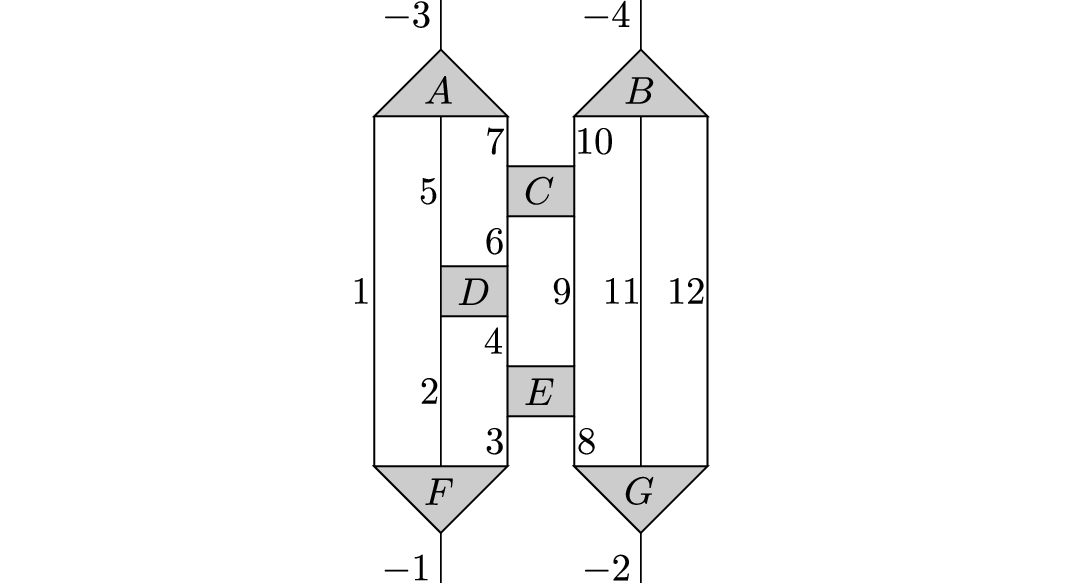}
\caption{\label{fig:labelled31MERA}A tensor network is described to \ttt{netcon()} using labelled indices. In this diagram,\arXtext{ which shows the same tensor network as \pfref{fig:31MERA},} summed indices have been labelled with unique positive integers, and open indices have been labelled with unique negative integers descending consecutively from $-1$. The tensors have also been labelled with letters to allow easy reference from the text.}
\end{figure}%
For each tensor~$T$, now construct a $1\times n_T$ matrix where $n_T$ is the number of indices attached to tensor~$T$, with entries corresponding to the labels associated with those indices. Ordering of the indices is unimportant. If there are $m$ tensors, then there are $m$ such matrices, which may be denoted $M_i,~i\in\{1\ldots m\}$.
Finally, \ttt{legLinks} comprises a $1\times m$ cell array, with the $m$ entries of this array corresponding to the matrices $M_i$, with ordering once again being unimportant. For example, the labelling given in \fref{fig:labelled31MERA} may be associated with the input parameter
\begin{align*}
\ttt{legLinks~=~\{}&\ttt{[-1 1 2 3],[2 4 5 6],[1 5 7 -3],}\\
&\ttt{[3 8 4 9],[6 9 7 10],[-2 8 11 12],}\\
&\ttt{[10 11 12 -4]\}}.
\end{align*}

\ttt{verbosity}: Determines the level of output generated by \ttt{netcon()}. For $\ttt{verbosity}=0$, an optimal index contraction sequence is returned in \ttt{sequence} and the associated cost is returned in \ttt{cost} but operation is otherwise silent. For $\ttt{verbosity}=1$, a message is generated every time the upper bound on tensor contraction costs is increased as described in \srefp{sec:costcap}, and on completion an optimal sequence and the associated cost are displayed on screen. For $\ttt{verbosity}=2$, behaviour is as for $\ttt{verbosity}=1$ but candidate contraction sequences and associated costs are announced when these sequences constitute the lowest-cost contraction sequence found so far. The final sequence and cost announced then correspond to the optimal solution reported at $\ttt{verbosity}=1$ and returned in the output variables \ttt{sequence} and \ttt{cost}. Setting $\ttt{verbosity}=3$ displays additional information about the pairwise contractions being performed. %

\ttt{costType}: To determine an optimal contraction sequence associated with a given tensor network, it is necessary to specify the dimension of each index in the network. Index dimensions may either be specified as integers, or in the form $a\chi^b$, where $\chi$ is an unspecified parameter which is presumed to be large. The former is indicated by $\ttt{costType}=1$, and the latter by $\ttt{costType}=2$. Default value: 2. Note that
when using $\ttt{costType}=2$, indices of some fixed dimension $d$ (such as the physical indices of an MPS or PEPS) may be represented by setting $a=d$ and $b=0$.

\ttt{muCap}: When searching for an optimal contraction sequence, \ttt{netcon()} initially restricts itself to sequences having a cost of at most $\xicap$ (for $\ttt{costType}=1$) or $\OO{\chi^\xicap}$ (for $\ttt{costType}=2$), where \ttt{muCap} represents $\xicap$. The value of \ttt{muCap} will automatically increase if no contraction sequence exists which satisfies this constraint. Setting \ttt{muCap} too high can incur extremely large overheads, whereas the process of automatic increase is relatively low-cost due to the caching of data in $S_i$ described in Secs.~\ref{sec:breadthalg} and \refp{sec:costcap}. It is therefore recommended that \ttt{muCap} be left at its default value of 1 unless the cost to contract the network is already known.

\ttt{allowOPs}: Determines whether \ttt{netcon()} should examine contraction sequences involving outer products (e.g.~$A^\alpha_\beta B^\gamma_\delta=C^{\alpha\gamma}_{\beta\delta}$). For tensor networks where there exists an optimal contraction sequence which does not involve an outer product, setting \ttt{allowOPs} to \ttt{false} may result in faster performance. However, if an outer product is required to obtain an optimal contraction sequence, this will also result in a suboptimal sequence being returned. Default value: \ttt{true}.

\ttt{legCosts}: The format of this input parameter is dependent upon the value of \ttt{costType}.
\begin{enumerate}
\item For $\ttt{costType}=1$, \ttt{legCosts} is an $\ell\times 2$ matrix whose first column consists of index labels and whose second column gives the dimensions associated with those labels. If \ttt{legCosts} is not specified, it is assumed that each index has dimension 2. 
\item For $\ttt{costType}=2$, \ttt{legCosts} comprises an $\ell\times 3$ matrix where $\ell$ is the total number of unique index labels. Each row then comprises three entries, $[x~a~b]$, where $x$ is a index label (and hence a positive or negative integer), and $a$ and $b$ specify the dimension of index $x$ in terms of the cost parameter $\chi$, such that $\mrm{dim}(x)=a\chi^b$. If \ttt{legCosts} is not specified, it is assumed that each index (whether summed or unsummed) has dimension $\chi$. Note that $b$ may take the value zero, permitting a fixed cost to be specified for some indices. 
\end{enumerate}
Regardless of the value of \ttt{costType}, if \ttt{legCosts} is specified, each index label must appear in the first column precisely once. Note that tensors of total dimension~1 are not supported.

On completion, \ttt{netcon()} returns an optimal contraction sequence and associated cost. These are specified as follows:

\ttt{sequence}: Sequence over which the indices of the tensor network should be summed in order to contract the network for minimum cost. For the interpretation of this sequence, see \condaref{sec:sequencenotation}.

\ttt{cost}: Specifies the total number of multiplication operations associated with optimal contraction of the tensor network, for example according to the sequence returned in \ttt{sequence}. For $\ttt{costType}=2$ this value is a number. For $\ttt{costType}=1$ the cost takes the form of a polynomial in $\chi$,
\begin{equation}
\sum_{i=0}^{\chi_\mrm{max}}a_i\chi^i,
\end{equation}
and this is returned as a $1\times \chi_\mrm{max}$ array whose entries \ttt{cost(i)} correspond to the coefficients $a_{i-1}$. %

Note that when a tensor network involves one or more traces, these may always be evaluated before network contraction begins. Evaluating a trace involves only addition operations, not multiplication, and thus these operations are relatively cheap and are ignored when computing the value of \ttt{cost} (though the presence of costs associated with tracing over indices will be noted in the text output if $\ttt{verbosity}>0$). %

\varsubsection{Index sequence notation\label{sec:sequencenotation}}

In the main body of this paper we have employed a heirarchical tensor-based notation to describe sequences of tensor contractions, with $((AB)C)$ indicating the contraction of tensor $A$ with tensor $B$ over all common indices (if any), followed by the contraction of the resulting object with tensor $C$. This notation is clear, unambiguous, and easily human-readable. In software implementations of tensor network algorithms, however, a linear notation is more widely used.

In this notation the indices of the tensor network are labelled as in \fref{fig:labelled31MERA}, and contractions are described by specifying the order in which the index sums are to be performed. For example, in the sequence
\begin{equation}
\ttt{[11 12 9 4 6 5 7 1 2 3 8 10]}
\end{equation}
the first sums are over indices 11 and 12, corresponding to contraction of tensor $B$ of \fref{fig:labelled31MERA} with tensor $G$. The next is over index 9, corresponding to contraction of tensor $C$ with tensor $E$. Proceeding in this fashion, the sequence is seen to describe the tensor contraction
\begin{equation}
((BG)((((CE)D)A)F)).
\end{equation}

Using an index-based notation it is also possible to describe sequences having no direct counterpart in the pairwise tensor contraction notation, for example
\begin{equation}
\ttt{[11 9 4 6 5 7 1 2 3 8 10 12]}.\label{eq:deferredtrace}
\end{equation}
In this contraction sequence, tensors $B$ and $G$ are combined by contracting over index~11 but index~12 remains unsummed until the end of the contraction sequence. In conventional tensor notation, writing $T$ for $(BG)$, one may write
\begin{equation}
T^{aef}_{bdg}=B^a_{bcd}G^{ecf}_g\label{eq:suboptimal}
\end{equation}
but the subsequent contraction over index~12 implies that we need only compute the diagonal elements
\begin{equation}
T^{aed}_{bdg}=B^a_{bcd}G^{ecd}_g,\qquad\textrm{no sum over $d$,}\label{eq:sparsetensor}
\end{equation}
similar to the calculation of $\left(C^*\right)^l_{ijkl}$ in \Erefp{eq:lunsummed}. However, as noted in \srefp{sec:flops} and in \rcite{lam1997}, it is never suboptimal to immediately sum such a repeated index. The behaviour of tensor contraction software for index sequences such as that in \Eref{eq:deferredtrace} may vary, with possibilities including producing a warning and performing the suboptimal contraction of \Eref{eq:suboptimal}, pre-emptive contraction of index~12 at the same time as index~11 regardless of the supplied sequence, or construction of a sparsely-populated tensor diagonal in index~12 as per \Eref{eq:sparsetensor}. To avoid this ambiguity, \ttt{netcon()} always returns sequences in which all index contractions between a given pair of tensors are performed at the same time. (An exception is made for indices of dimension one---the special handling of these indices is described in \condaref{sec:subnets}).

We also find it necessary to introduce an extension to the usual index sequence notation in order to describe tensor contraction sequences involving outer products. Although this notation is not universal---it cannot describe an entirely arbitrary contraction sequences involving outer products---it nevertheless suffices to describe all outer products consistent with the restrictions of \srefp{sec:restrictops}, and thus there always exists an optimal contraction sequence which can be expressed in this form.

In this extension, we use the label~0 to denote an outer product between two tensors, with a string of $n$ zeros denoting $n$ outer product operations between a total of $n+1$ tensors. These tensors are identified as follows:
\begin{itemize}
\item If there are $n$ zeros and only $n+1$ tensors remaining to contract, the outer product is between all remaining tensors.
\item Otherwise, read indices from the sequence following the zeros, and note the tensors they belong to, until $n+2$ tensors have been identified. By \scref{item:(AB)noopC} in \srefp{sec:tensorsinOP}, these $n+2$ tensors will comprise $n+1$ tensors participating in the outer product, and one tensor which shares indices with all of the other $n+1$ tensors. Perform the outer product on the $n+1$ tensors thus identified.
\end{itemize}
Outer products on multiple tensors should be performed pairwise, always acting on the two tensors of smallest total dimension.

An example tensor network requiring an outer product for optimal contraction is given in \fref{fig:disjointsequence}.
\begin{figure}
\includegraphics[width=246.0pt]{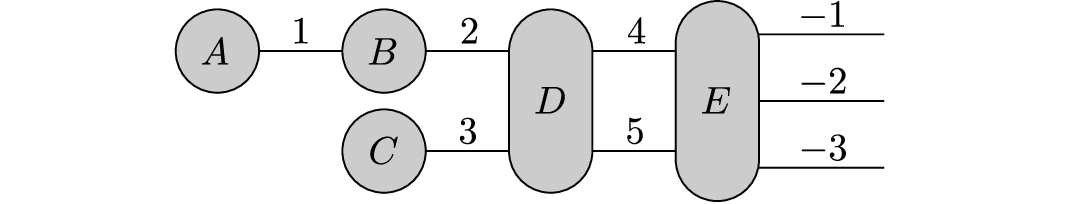}
\caption{An example tensor network for which the optimal contraction sequence involves performing an outer product. All indices are of dimension $\chi$, with $\chi>1$.\label{fig:disjointsequence}}
\end{figure}%
The optimal contraction sequence for this network, in index notation, is
\begin{equation*}
\ttt{[1 0 2 3 4 5]}
\end{equation*}
which corresponds to a pairwise tensor contraction sequence of
$((((AB)C)D)E)$.

A detailed algorithm for parsing index-based contraction sequences, including sequences which contain zeros to denote outer products, is provided for reference in \condaref{sec:readingsequence}. The syntax described, including the use of zeros to denote outer products, is fully supported by the tensor network contraction packages \ttt{ncon()} and \ttt{multienv()} of Refs.~\onlinecite{pfeifer2014} and~\onlinecite{evenbly2014} respectively.

\varsubsection{Disjoint subnetworks\label{sec:subnets}}

A tensor network is described as disjoint if it may be written as two or more factors sharing no common indices, e.g.~\Erefp{eq:disjoint}. Where a network is made up of multiple disjoint components, it is almost always preferable to address each component individually (with the only caveat relating to components which reduce to a scalar, as discussed in footnote~\refp{fn:scalar}). While, strictly speaking, the handling of such networks lies outside the scope of a reference implementation of the algorithm of \aref{sec:pseudocode}, support for disjoint networks has nevertheless been included in the interest of providing an implementation of maximal usefulness to the tensor network community. Each subnetwork is processed individually, and the resulting contraction sequences are concatenated together using outer products, represented by a string of zeros at the end of the contraction sequence.

As the computational cost of finding an optimal contraction sequence scales non-polynomially in $N$, the number of tensors, there is substantial benefit to be gained by identifying disjoint subnetworks and treating them independently. In light of this, it is noted that summing over an index of dimension one is formally equivalent to performing an outer product. For example, a column vector may be denoted either $A^a$ or $A^a_b$ where the only admissible value for index $b$ is 1, and the calculation
\begin{equation}
C^a_c = A^a_b B^b_c
\end{equation}
is entirely equivalent to
\begin{equation}
C^a_c = A^a B_c.
\end{equation}
Provided outer products between disconnected tensors are considered as part of a search algorithm, there is no disadvantage to ignoring or deleting all summed indices of dimension one in a tensor network. The reference implementation of \ttt{netcon()} ignores the existence of these indices both when identifying disjoint subnetworks and when determining optimal contraction sequences, and defers the zero-cost action of ``tracing over all indices of dimension one'' until the very end of the index contraction sequence. Because these indices are of dimension one there is never any cost penalty associated with doing so, even given the most pessimistic interpretation of contraction sequence notation as illustrated in \Eref{eq:suboptimal}.

\varsubsection{Sample invocation and output}

As a simple example, consider the tensor network given in \fref{fig:labelled31MERA}. Allowing \ttt{costType} and \ttt{legCosts} to take their default values, corresponding to each index having dimension $\chi$, \ttt{netcon()} may be invoked with verbosity level 1 by the command
\begin{align*}
\ttt{[sequence cost]=netcon(\{}&\ttt{[-1 1 2 3],[2 4 5 6],}\\
&\ttt{[1 5 7 -3],[3 8 4 9],}\\
&\ttt{[6 9 7 10],[-2 8 11 12],}\\
&\ttt{[10 11 12 -4]\},1);}
\end{align*}
with the pure-\MATLAB{} version returning the output
\begin{verbatim}
Looking for solutions of cost O(X^1)
Looking for solutions of cost O(X^6)
Looking for solutions of cost O(X^7)
Looking for solutions of cost O(X^8)
 
Best sequence:   11 12 9 4 6 5 7 1 2 3 8 10
Cost:            2X^8 + 2X^7 + 2X^6 + 0X^5
                 + 0X^4 + 0X^3 + 0X^2 + 0X^1
                 + 0X^0
\end{verbatim}
indicating that the cost of contracting this network scales as $\mrm{O}(\chi^8)$, and that for any given value of $\chi$ the actual cost of performing this contraction will be $2\chi^8+2\chi^7+2\chi^6$. The sequence and cost are also returned in the variables \ttt{sequence} and \ttt{cost}:
\begin{equation*}
\begin{tabular}{l}
\ttt{sequence = [11 12 9 4 6 5 7 1 2 3 8 10]}\\
\ttt{cost~~~~~= [0 0 0 0 0 0 2 2 2]}.
\end{tabular}
\end{equation*}
(Note that the pure-\MATLAB{} and \MATLAB{}-and-C++ versions use different methods of internally tabulating the results of intermediate contractions and so will frequently return different sequences to one another, but with the same contraction cost. This is because each version returns the first sequence of optimal cost which it encounters, and this may vary depending upon the order in which all relevant contractions are explored.)

\varsubsection{Using \ttt{netcon()} with \ttt{multienv()}}

The reference implementation of the ideas presented in this paper, \ttt{netcon()}, produces output compatible with the \ttt{multienv()} package described in \rcite{evenbly2014}. However, the networks supplied to \ttt{multienv()} are always \emph{closed} networks (i.e.~with no uncontracted indices), and we can take advantage of this. The \ttt{multienv()} package does not require an optimal contraction sequence for the supplied network, only a contraction sequence in the optimal \emph{family}, as defined in \rcite{evenbly2014}. This may be obtained at cheaper computational cost by deleting one tensor from the network, using \ttt{netcon()} to determine an optimal sequence for the open network, and finally reintroducing the deleted tensor, with contraction over the indices on this tensor being the final step in the contraction of the closed tensor network. A sequence obtained in this manner, while not necessarily optimal for contraction of the entire closed network, is nevertheless a member of the optimal family of sequences, and suffices for \ttt{multienv()} to calculate all requested tensor environments at minimal cost.

\section{Interpretation of contraction sequences specified as a list of indices\label{sec:readingsequence}}

The \ttt{netcon()} algorithm described in this paper takes as its input a description of a tensor network where each summed index is associated with a positive integer label and each open index is associated with a negative integer label. As its output the algorithm returns an optimal contraction sequence for the specified tensor network, specified as a list of
positive integer labels possibly interspersed with zeros,
and the cost of performing this contraction (corresponding to the number of multiplication operations required). 
Interpretation of tensor contraction sequences specified in this form is described in the paper above, but is summarised in this Appendix for convenience.

Starting with a list of tensors in the network to be contracted, and beginning with the first index of the sequence, contraction of a tensor network proceeds as follows:
\begin{description}[align=left,font=\normalfont,itemsep=2pt]
\item[1] If the sequence list is empty, stop. Contraction of the tensor network is complete.
\item[2] Read the first entry, $i_1$. 
\item[3] If $i_1=0$:
\begin{description}[align=left,font=\normalfont,itemsep=2pt]
\item[3a] Read a further $x-1$ entries, for a total of $x$ entries, denoted $i_1,\ldots,i_x$, where $x$ is the largest possible value such that all $x$ entries are zero.
\item[3b] Let $n$ be the number of tensors currently in the list of tensors. If $x=n-1$:
\begin{description}[align=left,font=\normalfont,itemsep=2pt]
\item[3b1] Using the outer product algorithm given below, perform an outer product of all $n$ remaining tensors. Denote the result of this outer product $X$. Delete all $n$ tensors from the list. Add tensor~$X$ to the list.
\item[3b2] Delete entries $i_1,\ldots,i_x$ from the sequence.
\item[3b3] Go to step 1.
\end{description}
\item[3c] Otherwise (i.e.~$x\not=n-1$):
\begin{description}[align=left,font=\normalfont,itemsep=2pt]
\item[3c1] Delete entries $i_1,\ldots,i_x$ from the sequence.
\item[3c2] Read the next $y$ entries from the sequence (denoted $j_1,\ldots,j_y$), and list all tensors on which indices $j_1,\ldots,j_y$ appear, where $y$ is the largest possible value such that all entries $j_1,\ldots,j_y$ are nonzero and the number of tensors in the list is precisely $x+2$.
\item[3c3] Let these tensors be referred to as ${A}_1,\ldots,{A}_{x+2}$.
\item[3c4] Let ${B}_1$ and ${B}_2$ denote the tensors on which index $j_1$ appears. Identify the smallest value of $z$ such that index $j_z$ appears on a tensor which is neither ${B}_1$ nor ${B}_2$. Index $j_z$ also appears on either ${B}_1$ or ${B}_2$. Let $X$ denote this tensor (either ${B}_1$ or ${B}_2$). Note that tensor~$X$ will be a member of the list ${A}_1,\ldots,{A}_{x+2}$.
\item[3c5] Using the outer product algorithm given below, perform an outer product of all tensors ${A}_1,\ldots,{A}_{x+2}$ except for tensor~$X$. Let the result of this outer product be denoted $Y$.
\item[3c6] Indices $j_1,\ldots,j_y$ all appear on both $X$ and $Y$. Evaluate the product of tensors~$X$ and $Y$, denoted $(XY)$, summing over all possible configurations of the indices $j_1,\ldots,j_y$.
\item[3c7] Delete tensors~${A}_1,\ldots,{A}_{x+2}$ from the list of tensors. Add tensor $(XY)$ to the list of tensors.
\item[3c8] Delete indices $j_1,\ldots,j_y$ from the sequence.
\item[3c9] Return to step 1.
\end{description}
\end{description}
\item[4] Otherwise (i.e.~$i_1\not=0$), identify which tensors index $i_1$ appears on.
\item[5] If index $i_1$ appears on only one tensor, it represents a trace. Read a further $x-1$ indices, for a total of $x$ indices, denoted $i_1,\ldots,i_x$, where $x$ is the largest possible value such that indices $i_1,\ldots,i_x$ are all traces and all appear on the same tensor. Trace over indices $i_1,\ldots,i_x$, delete these indices from the sequence, and return to step~1. (It is permissible for $x-1$ to be zero.)
\item[6] Otherwise: Index $i_1$ appears on two tensors, $A$ and $B$. Read a further $x-1$ indices, for a total of $x$ indices, denoted $i_1,\ldots,i_x$, where $x$ is the largest possible value such that all indices $i_1,\ldots,i_x$ appear on both tensor~$A$ and tensor~$B$. (It is permissible for $x-1$ to be zero.)
\item[7] Evaluate the product of tensors~$A$ and $B$, denoted $(AB)$, summing over all possible configurations of the indices $i_1,\ldots,i_x$.
\item[8] Delete tensors~$A$ and $B$ from the list of tensors. Add tensor $(AB)$ to the list of tensors.
\item[9] Delete indices $i_1,\ldots,i_x$ from the sequence.
\item[10] Return to step 1.
\end{description}

When called upon to perform an outer product of $m$ tensors, this should be done by applying the following algorithm:
\begin{description}[align=left,font=\normalfont]
\item[1] Define the total dimension of a tensor as the product of the dimensions of its indices.
\item[2] List the $m$ participating tensors.
\item[3] Identify the two tensors having the smallest total dimensions.
\item[4] Remove those two tensors from the list.
\item[5] Add their outer product to the list.
\item[6] Repeat steps 3-5 until only one tensor remains.
\end{description}
}
~

\section{Optimal contraction cost and index dimension\label{sec:chidependency}}

In \sref{sec:results} we computed contraction sequences for seven different tensor networks, expressing the costs of these sequences as polynomials in a parameter $\chi$ corresponding to index dimension. %
The costs and sequences computed are optimal for sufficiently large values of $\chi$, but it is important to note that the optimal contraction sequence may depend on the value of $\chi$ in a non-trivial fashion. As an example,
consider the simple network shown in \fref{fig:threetensors}.
Contracting this network according to the sequence $((AC)B)$ incurs a cost of $72\chi^6$, while the sequence $((BC)A)$ incurs a cost of $12\chi^7$. The sequence $((AC)B)$ is therefore clearly to be preferred in the limit of large $\chi$, but sequence $((BC)A)$ is preferred for index dimensions corresponding to values of $\chi<6$. When comparing the costs of different contraction sequences it is therefore necessary either to know the value of $\chi$, or to assume that $\chi$ is large. The reference implementation of\prtext{ \aref{sec:pseudocode} which is provided in the Supplementary Material \cite{EPAPS}}\arXtext{ Appendices~\ref{sec:pseudocode} and~\ref{sec:refimp}} permits either scenario, allowing index dimensions to be specified either as explicit real numbers, or as monomials $a\chi^b$ for real positive $a$ and real non-negative integer $b$. When index dimensions are specified as monomials, The value of $\chi$ is assumed to be large.

\bibliography{Netcon}

\end{document}